\documentclass[twocolumn]{IEEEtran}
\usepackage{cite}
\usepackage[cmex10]{amsmath}
\usepackage{array}
\usepackage{amssymb}
\usepackage{amsfonts}
\usepackage{subeqnarray}
\usepackage{cases}
\usepackage{mathrsfs}
\usepackage{graphicx}
\usepackage{threeparttable}
\usepackage{acronym}
\usepackage{caption}
\usepackage{subcaption}
\usepackage{dsfont}
\usepackage{url}
\usepackage[linesnumbered,ruled,vlined]{algorithm2e}
\newtheorem{theorem}{Theorem}

\newtheorem{definition}{Definition}
\newtheorem{example}{Example}

\newtheorem{lemma}[theorem]{Lemma}

\newtheorem{problem}{Problem}

\newtheorem{remark}{Remark}

\begin{document}

\title{Sensor placement by maximal projection on minimum eigenspace for linear inverse problems}

\author{\IEEEauthorblockN{Chaoyang~Jiang,~ Yeng~Chai~Soh, ~and~Hua~Li}

\thanks{Copyright (c) 2015 IEEE. Personal use of this material is permitted. However, permission to use this material for any other purposes must be obtained from the IEEE by sending a request to pubs-permissions@ieee.org.

The work was supported by Singapore's National Research Foundation under Grant NRF2011NRF-CRP001-090 Award Number NRF-CRP8-2011-03, and partially supported by the Energy Research Institute at NTU(ERI@N).

C. Jiang and Y. C. Soh are with the School of Electrical and Electronic Engineering, Nanyang Technological University, Singapore 639798 (e-mail: chaoyangjiang@hotmail.com; eycsoh@ntu.edu.sg).

H. Li is with the School of Mechanical and Aerospace Engineering, Nanyang Technological University, Singapore 639798 (e-mail: lihua@ntu.edu.sg.).

}
}

\maketitle

\begin{abstract}
This paper presents two new greedy sensor placement algorithms, named minimum nonzero eigenvalue pursuit (MNEP) and maximal projection on minimum eigenspace (MPME), for linear inverse problems, with greater emphasis on the MPME algorithm for performance comparison with existing approaches. In both MNEP and MPME, we select the sensing locations one-by-one. In this way, the least number of required sensor nodes can be determined by checking whether the estimation accuracy is satisfied after each sensing location is determined. For the MPME algorithm, the minimum eigenspace is defined as the eigenspace associated with the minimum eigenvalue of the dual observation matrix. For each sensing location, the projection of its observation vector onto the minimum eigenspace is shown to be monotonically decreasing w.r.t. the worst case error variance (WCEV) of the estimated parameters. We select the sensing location whose observation vector has the maximum projection onto the minimum eigenspace of the current dual observation matrix. The proposed MPME is shown to be one of the most computationally efficient algorithms. Our Monte-Carlo simulations showed that MPME outperforms the convex relaxation method \cite{joshi2009sensor}, the SparSenSe method \cite{jamali2014sparsity}, and the FrameSense method \cite{ranieri2014near} in terms of WCEV and the mean square error (MSE) of the estimated parameters, especially when the number of available sensor nodes is very limited.
\end{abstract}

\IEEEpeerreviewmaketitle

\section{Introduction}

\IEEEPARstart{S}{ensor} networks are widely used for monitoring temporal-spatial physical fields. While each sensor node can only observe the field intensity (e.g., temperature, humidity, concentration of contaminant, etc.) of a particular location, with a network of sparse sensor observations, a physical field of interest may be reconstructed by solving a linear inverse problem \cite{ranieri2014near, liu2014energy, cohen2006heuristic, willcox2006unsteady, Yildirim2009efficient, astrid2008missing}. In physical field estimation, the number of sensor nodes and their spatial locations are closely related to the coverage, cost, battery energy consumption, and even the error of the estimated physical field. Therefore, the determination of the least number of required sensor nodes and their locations is critical in sensor network design.

For a linear inverse problem, sensor placement is to seek the least number of required sensor nodes and their corresponding sensing locations within a known spatial domain such that the estimation accuracy can meet the requirement. Specifically, assuming that the observation models of all
potential sensing locations are known, we want to determine the least number of required sensors with which the physical field of interest can be recovered within a predefined accuracy. Obviously, one straightforward method is to evaluate the performance of all possible combinations of all potential sizes of the candidate sensing locations, and then select the one with the least number of sensor nodes that satisfies the required estimation accuracy. But such a combinatorial approach is computationally intractable.
In practice, direct enumeration is impossible if the number of potential sensing locations is large. Apart from the enumeration method, the optimal solution can also be obtained by branch-and-bound methods \cite{lawler1966branch, welch1982branch}, which unfortunately do take a very long time, even for a moderate scale problem \cite{joshi2009sensor}.
 Consequently, in recent years the sensor placement for the linear inverse problem has attracted increasing attention to find a suboptimal solution via computationally more efficient methods \cite{ranieri2014near, liu2014energy, Yildirim2009efficient, astrid2008missing, joshi2009sensor, jamali2014sparsity, chepuri2015continuous, shamaiah2010greedy, cohen2006heuristic, willcox2006unsteady, yao1993sensor, lau2008tabu, naeem2009cross}.

\subsection{Related prior work}
%Ranieri et al.  \cite{ranieri2014near} classified the solutions of this sensor placement problem into three categories: convex optimization, heuristics, and greedy methods, which will be discussed as follows.
Heuristics have been proposed to reduce the cost of exhaustive search.  The simplest one is to place sensor nodes at the spatial maxima and minima of proper orthogonal components of the physical field of interest \cite{cohen2006heuristic}. This method is simple but only suitable for some special cases \cite{willcox2006unsteady}. Other heuristics include genetic algorithms \cite{yao1993sensor}, particle swarm optimizer\cite{lau2008tabu}, tabu search \cite{lau2008tabu}, and cross-entropy optimization \cite{naeem2009cross}. They all involve a prohibitive computational cost and the solutions have no optimality guarantee.

Joshi and Boyd \cite{joshi2009sensor} formulated the sensor placement problem as an elegant nonconvex optimization problem, and approximated it as a convex optimization problem by the relaxation of the nonconvex Boolean constraints that represent the sensor placements,  to a convex box set. This convex relaxation was then used in many works \cite{chepuri2015sparsity, jamali2014sparsity, chepuri2013sparsity, chepuri2015continuous, mo2011sensor, shen2014sensor}.  The sensing locations can be easily determined based on the solution of the convex optimization problem. But the sensor placement may lead to an ill-conditioned observation model due to the gap between the nonconvex and the convex optimization problems, especially when the number of sensor nodes is very limited. Such a result has been shown to be no better than other works \cite{mo2011sensor, shen2014sensor, ranieri2014near}. However, the authors in \cite{joshi2009sensor} provided a local optimization technique to improve the result. This technique is computationally expensive but some numerical examples showed that with the local optimization, the convex relaxation method can indeed provide good results.

The sensor placement problem was also solved by some greedy algorithms in which the sensor locations are individually determined by optimizing some proxies of the error of the estimated physical field, such as the determinant of Fisher information matrix \cite{shamaiah2010greedy}, and the condition number \cite{Yildirim2009efficient, astrid2008missing, willcox2006unsteady} or the frame potential \cite{ranieri2014near} of the observation matrix. The $\eta$\emph{-confidence ellipsoid} of the estimation error depends on the determinant of the Fisher information matrix \cite{joshi2009sensor}, which was optimized using one greedy method in \cite{shamaiah2010greedy}, but it is shown to be no better than other methods in the examples in \cite{ranieri2014near}. For the sensor placement problem, the minimum requirement of the solution is that the observation model should be well-conditioned. Therefore, some researchers determined the sensing locations by minimizing the condition number of the observation matrix \cite{willcox2006unsteady, Yildirim2009efficient, astrid2008missing}. However, the condition number is a concept for nonsingular matrix, and we need to firstly determine a group of sensing locations to guarantee that the observation matrix is nonsingular \cite{Yildirim2009efficient}, which is unfortunately a combinatorial problem. Additionally, the minimum condition number of the observation matrix does not mean the minimum estimation error except when all the observation vectors have the same norm because the sensing energy should be considered, which is related to the signal-to-noise ratio. Recently, Ranieri et al. \cite{ranieri2014near} provided a novel greedy algorithm by minimizing the frame potential of the observation matrix. This method is computationally efficient but: 1) like the condition number minimization, it is only effective for the case where all the observation vectors have the same norm; 2) it cannot guarantee that the observation matrix is well-conditioned.

All the above mentioned works focused on the case where the number of sensor nodes is fixed. One sparse-promoting technique has been used to minimize the number of required sensor nodes by adding a sparsity-promoting penalty term to the cost function \cite{jamali2014sparsity}. This method works well when the dimension of the estimated parameter is small (e.g., the dimension is set as 2 in the example of Ref. [2]). However, if the dimension of the estimated parameter is large (e.g., a few tens, which is very common in fluid field reconstruction problems \cite{willcox2006unsteady, Yildirim2009efficient, iuliano2013proper}), this method will be ineffective in determining the least number of required sensor nodes, which will be discussed in detail later.

Besides the sensor placement for linear inverse problems, many other excellent sensor placement works have focused on the continuous system \cite{chepuri2015continuous}, nonlinear model \cite{chepuri2015sparsity}, energy saving \cite{mo2011sensor, liu2014energy}, state estimation for dynamic system \cite{mo2011sensor, shen2014sensorAES, shen2014sensor, liu2014optimal, chepuri2014sparsityFiltering}, and Gaussian process interpolation \cite{liu2014sparsityKriging, wang2004entropy, mackay1992information, guestrin2005near}.

\subsection{Our contributions}
In this paper, we propose a new greedy algorithm to minimize the number of required sensor nodes and determine their locations for the linear inverse problem such that the estimation error meets the requirement. We determine the sensing locations one-by-one until the estimation accuracy is satisfied by maximizing the projection of each observation vector onto the eigenspace of the minimum eigenvalue of the current dual observation matrix. It is shown that such a projection is monotonically decreasing w.r.t. the \emph{worst case error variance} (WCEV).

Compared with the state-of-the-art, the proposed greedy algorithm which we call the \emph{maximal projection on minimum eigenspace} (MPME), has the following advantages:
\begin{itemize}
  \item The MPME can readily determine the minimum number of required sensor nodes.
  \item The MPME outperforms the convex relaxation method \cite{joshi2009sensor}, the SparSenSe method \cite{jamali2014sparsity}, and the FrameSense method \cite{ranieri2014near} in terms of the WCEV and the mean square error (MSE) of the estimated vector, especially when the number of available sensor nodes is very limited.
  \item The MPME can guarantee that the observation matrix is well-conditioned but the convex relaxation, the SparSenSe and the FrameSense methods cannot guarantee such a condition, especially when the number of available sensor nodes is very limited.
  \item For general sensor placement problems, the MPME without local optimization \cite{joshi2009sensor} outperforms the state-of-the-art with local optimization.
  \item The proposed MPME is computationally one of the most efficient sensor placement algorithms.
\end{itemize}

\subsection{Outline and notations}
The rest of this paper is organized as follows. In Section II, we introduce the linear inverse problem and the sensor placement problem, and briefly review three current methods. In Section III, we develop the MPME algorithm. We then provide four examples to compare the effectiveness of MPME with the current methods via Monte-Carlo simulations in Section IV. In Section V, we analyze the computational cost of the MPME algorithm and compare it with those of the current methods. The conclusions are given in Section VI.

This paper uses the following notations: Upper (lower) bold letters, e.g. $\mathbf{A}$ ($\mathbf{a}$) or $\mathbf{\Phi}$ ($\boldsymbol{\varphi}$), indicate matrices (column vectors). $\mathbf{I}$ represents an identify matrix with proper dimension whose $i$-th column vector is denoted by $\mathbf{e}_i$. $\mathbf{1}$ is a vector of proper dimension with all entries one.
$(\cdot)^\mathrm{T}$, $(\cdot)^\dagger$, $\mathbb{{E}}(\cdot)$, $\mathbf{tr}(\cdot)$, $\|\cdot\|$, $\mathbf{det}(\cdot)$, $\mathbf{span}(\cdot)$, $\mathbf{null}(\cdot)$, $\mathbf{dim}(\cdot)$, and $\mathbf{rank}(\cdot)$ are respectively the transposition, pseudo-inverse, expectation, trace, norm, determinant, spanned space, null space, dimension, and rank operators.

\section{Problem Statement}

\subsection{Linear inverse problem}
We consider a physical field $\mathbf{f}\in \mathbb{R}^N$ described as
\begin{equation}
\mathbf{f}=\tilde{\mathbf{\Phi}} \boldsymbol{\alpha} \label{eq:physicalField}
\end{equation}
where $ \boldsymbol{\alpha} \in \mathbb{R}^n$ is a vector of parameters to be estimated with $n\ll N$, and $\tilde{\mathbf{\Phi}}\in \mathbb{R}^{N\times n}$ is a known full column-rank matrix, which we call the \emph{signal representation matrix} and its column vectors compose a basis of the physical field.

It is expensive and impractical to sense the physical field $\mathbf{f}$ with $N$ sensor nodes since $N$ is very large and depends on the resolution of the discrete physical space \cite{ranieri2014near}. However, part of the physical field can be observed from sensor networks, i.e.
\begin{equation}
\mathbf{y}=\mathbf{H}\mathbf{f}+\boldsymbol\nu={\mathbf{\Phi}} \boldsymbol{\alpha}+\boldsymbol\nu \label{eq:LinearMeasurementModel}
\end{equation}
where $\mathbf{H}\in \mathbb{R}^{M\times N}$ whose $i$-th row is $\mathbf{e}_{s_i}^\mathrm{T}$, $s_i\in \mathcal{N}=\{1,2,...,N\}$ corresponds to the $i$-th sensing location, and $M (n\leq M \ll N)$ is the number of sensor nodes. The \emph{observation matrix}
\begin{equation*}
{\mathbf{\Phi}}=\mathbf{H}\tilde{\mathbf{\Phi}}=[\boldsymbol{\varphi}_{s_1}, \boldsymbol{\varphi}_{s_2},...\boldsymbol{\varphi}_{s_M}]^\mathrm{T}
\end{equation*}
is a pruned matrix from the rows of $\tilde{\mathbf{\Phi}}$ indexed by $\{s_1, s_2,..., s_M\}$, and $\boldsymbol{\varphi}_{s_i}^\mathrm{T}$ is the $s_i$-th row of $\tilde{\mathbf{\Phi}}$ and represents the observation model of the $i$-th sensor node, which we call the \emph{observation vector}. The measurement noise $\boldsymbol\nu\in \mathbb{R}^M$ is assumed to be zero-mean i.i.d. Gaussian random process with variance $\sigma^2\mathbf{I}$.

From \eqref{eq:LinearMeasurementModel}, we can obtain the following \emph{minimum variance unbiased estimate} (MVUE)
\begin{equation}
\hat{\boldsymbol{\alpha}} = {\mathbf{\Phi}}^{\dagger}\mathbf{y} \label{eq:MVUE}
\end{equation}
where ${\mathbf{\Phi}}^{\dagger}=({\mathbf{\Phi}}^\mathrm{T}{\mathbf{\Phi}})^{-1}{\mathbf{\Phi}}^\mathrm{T}$
is the pseudo-inverse of ${\mathbf{\Phi}}$. The mean square error (MSE) of this MVUE \cite{ranieri2014near, joshi2009sensor, jamali2014sparsity} is
\begin{equation}
\mathrm{MSE}(\hat{\boldsymbol{\alpha}}) = \mathds{E}\left( \|\hat{\boldsymbol{\alpha}}- {\boldsymbol{\alpha}}\|_2^2\right)=\sigma^2\mathrm{\mathbf{tr}} (\mathbf{\Psi}^{-1}) =\sigma^2\sum_{k=1}^n  \frac{1}{\lambda_k} \label{eq:MSE}
\end{equation}
where $\lambda_1\geq\lambda_2\geq...\geq\lambda_n$ stand for the eigenvalues of
\begin{equation*}
\mathbf{\Psi} = {\mathbf{\Phi}}^\mathrm{T}{\mathbf{\Phi}}
\end{equation*}
which we call the \emph{dual observation matrix}.

With some standard operations, we can obtain the variance of $\hat{\boldsymbol{\alpha}}$ as
\begin{eqnarray*}
\mathbf{\Sigma} &=& \mathds{E}\left[ (\hat{\boldsymbol{\alpha}} - {\boldsymbol{\alpha}})(\hat{\boldsymbol{\alpha}} - {\boldsymbol{\alpha}})^\mathrm{T}\right]\notag\\ &=&\mathds{E}\left[{\mathbf{\Phi}}^{\dagger}\boldsymbol\nu \boldsymbol{\nu}^\mathrm{T}({\mathbf{\Phi}}^{\dagger})^\mathrm{T}\right]\notag \\
&=& \sigma^2{\mathbf{\Phi}}^{\dagger}({\mathbf{\Phi}}^{\dagger})^\mathrm{T} \label{eq:Variance1}=\sigma^2 \mathbf{\Psi}^{-1}\label{eq:Variance2}
\end{eqnarray*}
Then, we introduce the following \emph{worst case error variance} (WCEV) of the MVUE $\hat{\boldsymbol{\alpha}}$
\begin{equation}
\mathrm{WCEV}(\hat{\boldsymbol{\alpha}} )= \underset{\|\mathbf{x}\|_2=1} {\mathrm{max}} ~\mathbf{x}^\mathrm{T}\mathbf{\Sigma}\mathbf{x}=\lambda_{\mathrm{max}}(\mathbf{\Sigma})=\frac{\sigma^2}{\lambda_n} \label{eq:WCEV}
\end{equation}
For more detail about WCEV, do refer to \cite{joshi2009sensor}. %Both $\mathrm{MSE}(\hat{\mathbf{a}})$ and $\mathrm{WCEV}(\hat{\mathbf{a}})$ are indicators to access the performance of the MVUE $\hat{\mathbf{a}}$.
Since ${\mathbf{\Phi}}^{\dagger}({\mathbf{\Phi}}^{\dagger})^\mathrm{T}$ $=\mathbf{\Psi}^{-1}$, it is easily found from \eqref{eq:MSE} and \eqref{eq:WCEV} that
\begin{eqnarray*}
\mathrm{MSE}(\hat{\boldsymbol{\alpha}})&=&\sigma^2\|{\mathbf{\Phi}}^{\dagger}\|_\mathrm{F}^2\\
\mathrm{WCEV}(\hat{\boldsymbol{\alpha}})&=&\sigma^2\|{\mathbf{\Phi}}^{\dagger}\|_2^2
\end{eqnarray*}
Consequently, the two error indicators are equivalent due to the \emph{equivalence} of the two matrix norms \cite{golub2012matrix}. Specifically,
\begin{equation}
\mathrm{WCEV}(\hat{\boldsymbol{\alpha}})\leq \mathrm{MSE}(\hat{\boldsymbol{\alpha}})\leq n\mathrm{WCEV}(\hat{\boldsymbol{\alpha}}) \label{eq:equivalence}
\end{equation}

It is clear in \eqref{eq:MSE} and \eqref{eq:WCEV} that both MSE and WCEV depend on the eigenvalues of the dual observation matrix $\mathbf{\Psi}$, which fully depends on the sensing locations.

\subsection{Sensor placement problem}
We denote the set of selected sensing locations by $\mathcal{S}=\{s_1, s_2,..., s_M\}$, and the set of potential sensing locations by $\mathcal{N}=\{1, 2,..., N\}$, which correspond to the row indices of $\mathbf{\Phi}$ and $\tilde{\mathbf{\Phi}}$, respectively. Then, we formulate the following sensor placement problem.

\begin{problem}
Given the signal representation matrix $\tilde{\mathbf{\Phi}}= [\boldsymbol\varphi_1, \boldsymbol\varphi_2,..., \boldsymbol\varphi_N]^\mathrm{T} \in \mathbb{R}^{N\times n}$, select $M$ rows of $\tilde{\mathbf{\Phi}}$ indexed by $s_1, s_2,..., s_M$ to construct the observation matrix ${\mathbf{\Phi}} = [\boldsymbol\varphi_{s_1}, \boldsymbol\varphi_{s_2},..., \boldsymbol\varphi_{s_M}]^\mathrm{T} \in \mathbb{R}^{M\times n}$, such that the error of the estimated parameters $\hat{\boldsymbol{\alpha}}$ in \eqref{eq:MVUE} is small enough and the number of rows of ${\mathbf{\Phi}}$, i.e. $M$, is minimized.
\end{problem}

This actually is a sensing location selection problem. We aim to find the minimum number of sensing locations with which the error of $\hat{\boldsymbol{\alpha}}$ is less than a predefined threshold. In this paper, we use the WCEV as the error indicator, and equation \eqref{eq:equivalence} shows that a small WCEV can guarantee a small MSE. Then, this sensor placement problem can be formulated as the following cardinality minimization problem
\begin{equation} \label{eq:WCEVOptimization}
\hat{\mathcal{S}}= \underset{\mathcal{S}\subseteq \mathcal{N}} {\arg\,\min} ~ |\mathcal{S}| \quad \mathrm{subject}~\mathrm{to} \quad \lambda_n\geq \gamma
\end{equation}
where $|\cdot|$ returns the cardinality of a set, and $\gamma$ corresponds to the maximum acceptable WCEV.

\subsection{The state of the art}
The combinatorial optimization problem \eqref{eq:WCEVOptimization} is NP-hard \cite{papadimitriou1998combinatorial}. Here, we briefly review three current and related methods. Two of them are originally designed for the case where the number of available sensors is fixed. However, they can be simply modified and applied to the case where the number of sensors is unknown, which is discussed in Remark 1.

\subsubsection{Convex relaxation \cite{joshi2009sensor}} If the number of sensor nodes is fixed, the sensor placement problem can be formulated as
\begin{eqnarray}
\mathrm{maximize} && \mathbf{{log}}  \; \mathbf{{det}}\left(\sum_{i=1}^Nw_i\boldsymbol\varphi_i\boldsymbol\varphi_i^\mathrm{T}\right) \notag\\
\mathrm{subject\,to}&& \mathbf{1}^\mathrm{T}\mathbf{w}=M\notag\\
&&w_i\in\{0,1\},\quad i\in \mathcal{N} \label{eq:nonconvexConstraint}
\end{eqnarray}
with variable $\mathbf{w}\in \mathbb{R}^N$. Here $w_i=1$ means $i\in \mathcal{S}$， and $w_i=0$ means $i\notin \mathcal{S}$. Performing a convex relaxation, i.e. replacing the nonconvex Boolean constraints $w_i\in\{0, 1\}$  by $w_i\in[0, 1]$, we can obtain the following convex optimization problem:
\begin{eqnarray}
\mathrm{maximize} && \mathbf{{log}}  \; \mathbf{{det}}\left(\sum_{i=1}^Nw_i\boldsymbol\varphi_i\boldsymbol\varphi_i^\mathrm{T}\right) \notag\\
\mathrm{subject\,to}&& \mathbf{1}^\mathrm{T}\mathbf{w}=M\notag\\
&&w_i\in[0,1],\quad i\in \mathcal{N} \label{eq:convexRelaxation}
\end{eqnarray}
with variable $\mathbf{w}$. This problem can be solved by the interior-point methods \cite{boyd2004convex}. Rearranging the entries of the solution of the relaxed problem \eqref{eq:convexRelaxation}, i.e. $\mathbf{w^*}$($\in[0, 1]^N$), in descending order yields the sequence $\{w_{\hat{s}_1}^*, w_{\hat s_2}^*,\dots, w_{\hat s_N}^*\}$. Then, the set of the sensing indices is given by $\hat{\mathcal{S}}=\{{\hat{s}_1}, {\hat s_2},\dots, {\hat s_M}\}$, i.e. the indices of the $M$ largest elements of $\mathbf{w}^*$.

\subsubsection{SparSenSe \cite{jamali2014sparsity}} To determine the number of required sensor nodes, the following convex optimization, called sparse-aware sensor selection (SparSenSe), is formulated:
\begin{eqnarray}
\mathrm{minimize} && \|\mathbf{w}\|_{l1} \notag\\
\mathrm{subject\,to}&& \left[\begin{array}{cc}
\sum_{i=1}^Nw_i\boldsymbol\varphi_i\boldsymbol\varphi_i^\mathrm{T} & \mathbf{e}_j\\ \mathbf{e}_j^\mathrm{T} & x_j
\end{array}\right]\succeq \mathbf{0},\; j=1,...,n\notag\\
&&\|\mathbf{x}\|_{l1}\leq \gamma',\; x_j \geq 0,  \; j=1,...,n\notag\\
&&w_i\in[0,1],\quad i\in \mathcal{N} \label{eq:SparSenSe}
\end{eqnarray}
where $\gamma'$ corresponds to the maximum acceptable MSE index. This is a linear matrix inequalities problem and can be solved by using the CVX toolbox \cite{Boydcvx}. With the solution $\mathbf{w}^*$ and \emph{a prior} threshold $\tau$ ($0\lessapprox\tau\ll1$), we can determine the sensing indices. If $w_i^* <\tau$, set $w_i^*=0$. The number of required sensor nodes is the number of nonzero entries of $\mathbf{w}^*$.

\subsubsection{FrameSense \cite{ranieri2014near}}
The ensemble of the rows of a matrix can be viewed as a frame. If all the observation vectors have the same norm, according to the frame theory, the observation matrix $\mathbf{\Phi}$ achieves the minimum MSE when it achieves the minimum frame potential \cite{ranieri2014near, benedetto2003finite}. For the basic concept of the frame theory, do refer to \cite{kovacevic2007life}. The sensor placement problem can be solved by minimizing the following frame potential

\begin{equation}
\mathrm{FP}(\mathbf{\Phi})=\underset{i,j\in \mathcal{S}}{\sum}(\boldsymbol\varphi_i^\mathrm{T}\boldsymbol\varphi_j)^2 \label{eq:FramePotential}
\end{equation}

One greedy ``worst-out" algorithm, called the FrameSense, can provide a near-optimal solution in the sense of the minimum frame potential. At each step, it removes the row of $\tilde{\mathbf{\Phi}}$ that maximally increases the frame potential. If $\tilde{\mathbf{\Phi}}$ corresponds to an equal-norm frame, the row index is in fact the index of the row/column of $(\tilde{\mathbf{\Phi}}\tilde{\mathbf{\Phi}}^\mathrm{T})^2$  which has the largest 1-norm. Here, $\mathbf{A}^2$ denotes a matrix whose entries are the square of the corresponding entries of $\mathbf{A}$.

\begin{remark}
With simple modifications, convex relaxation and FrameSense can be used to determine the least number of required sensors. For the convex relaxation method, it can be found by increasing the sensor number from $n$ until the constraint in \eqref{eq:WCEVOptimization} is satisfied. For FrameSense, when removing each row of $\tilde{\mathbf{\Phi}}$, we check the constraint in \eqref{eq:WCEVOptimization}. If the constraint is not satisfied, reserve the row and the number of remaining rows of $\tilde{\mathbf{\Phi}}$ is the least number of required sensors.
\end{remark}

\section{Maximal Projection on Minimum Eigenspace}

As mentioned before, one apparent method to solve the cardinality optimization problem \eqref{eq:WCEVOptimization} is to evaluate the minimum eigenvalue of the dual observation matrix (i.e. $\lambda_n$) of all potential sensor configurations, and then find the configuration with the minimum number of sensors that satisfies the constraint. Unfortunately, the computational cost of exhaustively searching $2^N$ potential configurations is unaffordable for large scale problems. One simple strategy to reduce the number of searched sensor configurations is to determine the sensing locations one-by-one. With such a strategy, the minimum number of required sensor nodes, $M$, can be easily found by judging whether the constraint in \eqref{eq:WCEVOptimization} is satisfied after each sensing location is determined. In this way, the number of searched sensor configurations can be reduced to $\sum_{i=0}^{M-1}(N-i)$, since at the first step over $N$ possible sensing locations are searched, then $N-1$, and so on.

Admittedly, each sensor node may have correlated influence with others, and one sensor reading is informative for a given sensor configuration but may be meaningless for others.
When finding the sensing locations one-by-one, we do not know the contribution of each sensor node for the final sensor configuration; therefore, such a strategy cannot guarantee the optimal solution. However, we are trying to make a tradeoff between the computational cost and the number of required sensor nodes, i.e. to find an effective sensor configuration with proper number of sensor nodes by determining the sensing locations one-by-one.

When determining the sensing locations one-by-one, we can obtain an observation vector sequence $\{\boldsymbol{\varphi}_{s_k}\}_{k=1}^M$. For simplicity, we introduce a new matrix $\mathbf{\Phi}_k\in \mathbb{R}^{k\times n}$ to denote the first $k (1\leq k \leq M)$ rows of $\mathbf{\Phi}$, i.e. $\mathbf{\Phi}_k=[\boldsymbol{\varphi}_{s_1}, \boldsymbol{\varphi}_{s_2},...\boldsymbol{\varphi}_{s_k}]^\mathrm{T}$ corresponds to the first $k$ sensing locations. Corresponding to $\{\boldsymbol{\varphi}_{s_k}\}_{k=1}^M$, we can obtain the matrix sequence $\{\mathbf{\Phi}_k\}_{k=1}^M$ and the dual observation matrix sequence $\{\mathbf{\Psi}_k\}_{k=1}^M$.
Here $\mathbf{\Psi}_k = \mathbf{\Phi}_k^\mathrm{T}\mathbf{\Phi}_k$ has a nonincreasing eigenvalue sequence $\{\lambda_i^{(k)}\}_{i=1}^n$.  The notations that will commonly appear in this paper are listed below for easy reference.

\begin{table}[ht]
\begin{center}
\begin{tabular}{ll}
\hline

{$\boldsymbol{\varphi}_{s_k}$} & the $k$-th selected observation vector corresponding to the $k$-th \\
&sensing location. $\boldsymbol{\varphi}_{s_k}^\mathrm{T}$ is the $k$-th row of $\mathbf{\Phi}$ and $s_k$-th row of $\tilde{\mathbf{\Phi}}$.\\

{$\mathbf{\Phi}_k$} &includes the first $k$ observation vectors. It consists of the first $k$ \\
&rows of the observation matrix $\mathbf{\Phi}=\mathbf{\Phi}_M$.\\

 {$\mathbf{\Psi}_k$} & $=\mathbf{\Phi}_k^\mathrm{T}\mathbf{\Phi}_k$, is the dual observation matrix associated with $\mathbf{\Phi}_k$.\\
{$\lambda_i^{(k)}$} & the $i$-th eigenvalue of $\mathbf{\Psi}_k$, i.e. $\lambda_i (\mathbf{\Psi}_k)$.\\
$\mathbf{u}_i^{(k)}$ &  the normalized eigenvector of $\mathbf{\Psi}_k$ associated with $\lambda_i^{(k)}$.\\
$\mu_n$ &  the multiplicity of the minimum eigenvalue $\lambda_n^{(k-1)}$ w.r.t. $\mathbf{\Psi}_{k-1}$.\\
{$\lambda_n$} & the minimum eigenvalue of $\mathbf{\Psi}=\mathbf{\Psi}_M$, i.e. $\lambda_n^{(M)}$.\\

\hline
\end{tabular}
\end{center}
\end{table}

For all $k<n$, the minimum eigenvalue of $\mathbf{\Psi}_k$, $\lambda_n^{(k)}=0$ because $\mathbf{rank}(\mathbf{\Psi}_k)=\mathbf{rank}(\mathbf{\Phi}_k)\leq k$.  Therefore, the number of required sensor nodes $M$ must be no less than $n$. Since $\tilde{\mathbf{\Phi}}$ is a full column-rank matrix, the nonsingular $\mathbf{\Psi}_n$ exists, which implies that the optimal $M$ may be equal to $n$. Consequently, to minimize $M$ we should guarantee that $\mathbf{\Psi}_n$ is nonsingular. In that case, the vectors in $\{\boldsymbol{\varphi}_{s_k}\}_{k=1}^n$ are mutually independent and $\mathbf{rank}(\mathbf{\Psi}_k)=\mathbf{rank}(\mathbf{\Phi}_k)=\mathbf{rank}(\{\boldsymbol{\varphi}_{s_i}\}_{i=1}^k)=k$ for all $k\leq n$.
In practice, we can easily guarantee that $\boldsymbol{\varphi}_{s_k}$ is independent with the vectors in $\{\boldsymbol{\varphi}_{s_i}\}_{i=1}^{k-1}$ when determining the $k$-th sensing location for all $k\leq n$.

Our purpose is to find the shortest observation vector sequence $\{\boldsymbol{\varphi}_{s_k}\}_{k=1}^M$ by determining the sensing locations one-by-one, such that the constraint in \eqref{eq:WCEVOptimization}, i.e. $\lambda_n\geq \gamma$, is satisfied.
For a given sensor configuration corresponding to $\mathbf{\Phi}_{k-1} (1\leq k< M)$, we need to formulate some guidelines to determine the next sensing location $s_{k}$. Intuitively, we can traverse all $N-k+1$ unselected observation vectors to find the one that can maximally increase $\lambda_n$. Then, the other sensing locations can be similarly determined one-by-one until $\lambda_n>\gamma$ and meanwhile the number of selected sensing locations is the minimum number of required sensor nodes, i.e. $M$.

However, the challenge is that we do not know how $\boldsymbol{\varphi}_{s_{k}}$ affects $\lambda_n(=\lambda_n^{(M)})$ since $M$ and $\{\boldsymbol{\varphi}_{s_{i}}\}_{i=k+1}^M$ are unknown when we find the $k$-th sensing location. In other words, we cannot build an explicit mapping between $\boldsymbol{\varphi}_{s_{k}}$ and $\lambda_n$. Therefore, it is impossible to find the optimal $k$-th sensing location $s_{k}^*$ by directly optimizing $\lambda_n$.

Our main idea is to find a new criterion instead of $\lambda_n$. When determining the $k$-th sensing location, we optimize the new criterion. The criterion should satisfy three conditions:
\begin{enumerate}
  \item It can be directly obtained from $\mathbf{\Phi}_{k-1}$ and $\boldsymbol{\varphi}_{s_{k}}$.
  \item Optimizing the criterion can guarantee that all the vectors in $\{\boldsymbol{\varphi}_{s_k}\}_{k=1}^n$ are mutually independent.
  \item $\lambda_n$ is positively correlated with the criterion.
\end{enumerate}
The first condition implies that we can directly assess the contributions of all potential observation vectors for the new criterion. The second condition guarantees that $\mathbf{\Psi}_n$ is nonsingular, and the last condition guarantees that increasing the new criterion can increase $\lambda_n$, i.e. decrease the WCEV.
If such a criterion exists, we can determine the sensing locations one-by-one by maximizing the new criterion. Accordingly, we can find a suboptimal sensor configuration. In what follows, we shall present two alternative criteria.

\subsection{Minimum nonzero eigenvalue pursuit (MNEP)} \label{Section:MNEP}

%If we determine the sensing locations one-by-one, corresponding to the observation vector sequence $\{\boldsymbol{\varphi}_{s_k}\}_{k=1}^M$, we can obtain the dual observation matrix sequence $\{\mathbf{\Psi}_k\}_{k=1}^M$.
It can be easily found that
\begin{equation*}
\mathbf{\Psi}_{k}=\mathbf{\Phi}_k^\mathrm{T}\mathbf{\Phi}_k=[\mathbf{\Phi}_{k-1}^\mathrm{T} \;\; \boldsymbol{\varphi}_{s_{k}}][\mathbf{\Phi}_{k-1}^\mathrm{T} \;\; \boldsymbol{\varphi}_{s_{k}}]^\mathrm{T}=\mathbf{\Psi}_{k-1} + \boldsymbol{\varphi}_{s_{k}}\boldsymbol{\varphi}_{s_{k}}^\mathrm{T}
\end{equation*}
This equation implies that all the eigenvalues of $\mathbf{\Psi}_k$ satisfy the first condition of the criterion used to replace $\lambda_n$, i.e. the eigenvalues of $\mathbf{\Psi}_k$ can be found if $\mathbf{\Phi}_{k-1}$ and $\boldsymbol{\varphi}_{s_{k}}$ are known.
Amongst all the eigenvalues of $\mathbf{\Psi}_k$, we guess that the \emph{minimum nonzero eigenvalue}, i.e. $\lambda_k^{(k)}$ when $k\leq n$ and $\lambda_n^{(k)}$ when $k>n$, is one choice of the criterion to be used to replace $\lambda_n$.

Obviously, for any $k\leq n$, maximizing $\lambda_k^{(k)}$ can guarantee that $\lambda_k^{(k)}>0$, which implies that for any $k\leq n$, the vectors in $\{\boldsymbol{\varphi}_{s_{i}}\}_{i=1}^{k}$ are mutually independent. Therefore, the minimum nonzero eigenvalue of $\mathbf{\Psi}_k$ satisfies the second condition. We then utilize the following theorem to show that it also satisfies the third condition.

\begin{theorem} \label{Theorem:sequence}
Suppose $\mathbf{B}=\mathbf{A}+\mathbf{c}\mathbf{c}^\mathrm{T}$ where $\mathbf{A}\in \mathbb{R}^{n\times n}$ is symmetric, and $\mathbf{c}\in \mathbb{R}^n$ is a non-zero vector. Then,
\begin{equation*}
\lambda_{1}(\mathbf{B}) \geq \lambda_1(\mathbf{A})\geq \lambda_{2}(\mathbf{B})\geq \lambda_{2}(\mathbf{A})\geq ...\geq \lambda_n(\mathbf{B})\geq \lambda_n(\mathbf{A})
\end{equation*}
\end{theorem}
\begin{IEEEproof}
See \cite{thompson1976behavior} and the Theorem 8.1.8 in \cite{golub2012matrix}.
\end{IEEEproof}

Since $\mathbf{\Psi}_{k}=\mathbf{\Psi}_{k-1} + \boldsymbol{\varphi}_{s_{k}}\boldsymbol{\varphi}_{s_{k}}^\mathrm{T}$, considering \emph{Theorem \ref{Theorem:sequence}}, we can obtain
\begin{eqnarray*}
&&\lambda_n^{(k)}\geq \lambda_{n}^{(k-1)} \quad \mathrm{for \;\;all} \quad k\geq n \\
&&\lambda_k^{(k)}\leq \lambda_{k-1}^{(k-1)}\quad \mathrm{for \;\;all}\quad k\leq n
\end{eqnarray*}
From the two equations, we can easily find that
\begin{eqnarray}
&&\lambda_n^{(M)}\geq \lambda_{n}^{(M-1)}\geq...\geq \lambda_n^{(n)} \label{eq:lambda_nk}\\
&&\lambda_n^{(n)}\leq \lambda_{n-1}^{(n-1)}\leq...\leq \lambda_1^{(1)}  \label{eq:lambda_kk}
\end{eqnarray}

For any $n\leq k< M$, \eqref{eq:lambda_nk} shows that $\lambda_n^{(k)}$ is the lower bound of $\lambda_n^{(k+1)}$.
If we maximize $\lambda_n^{(k)}$ by proper selection of $\boldsymbol{\varphi}_{s_k}$, we maximize the lower bound of $\lambda_n^{(k+1)}$. Hence, $\lambda_n^{(k+1)}$ is positively correlated with $\lambda_n^{(k)}$. Since $\lambda_n=\lambda_n^{(M)}\geq \lambda_n^{(M-1)}\geq...\geq \lambda_{n}^{(k+1)}\geq \lambda_n^{(k)}$, $\lambda_n$ is positively correlated with $\lambda_n^{(k)}$ for any $n\leq k <M$.

For any $k< n$, \eqref{eq:lambda_kk} shows that $\lambda_{k}^{(k)}$ is the upper bound of $\lambda_{k+1}^{(k+1)}$. Therefore, if we select sensing location to maximize $\lambda_k^{(k)}$, we maximize the upper bound of $\lambda_n^{(n)}$. Actually, $\lambda_{k+1}^{(k+1)}$ is monotonically increasing w.r.t. $\lambda_{k}^{(k)}$ for any $k<n$, which will be shown later. Hence, $\lambda_n^{(n)}$ is monotonically increasing w.r.t. $\lambda_{k}^{(k)}$ for all $k<n$.
Since $\lambda_n=\lambda_n^{(M)}$ is positively correlated with $\lambda_n^{(n)}$, we conclude that $\lambda_n$ is also positively correlated with $\lambda_k^{(k)}$ for any $k<n$.

In summary, $\lambda_n$ is positively correlated with the minimum nonzero eigenvalue of $\mathbf{\Psi}_k$, i.e. $\lambda_k^{(k)}$ for $k<n$ and $\lambda_n^{(k)}$ for $k\leq n$, which satisfies the third condition.

Therefore, the minimum nonzero eigenvalue of $\mathbf{\Psi}_k$ can be a criteria in place of $\lambda_n$ to optimize the $k$-th sensing location. To maximize $\lambda_n$, we can select $\boldsymbol{\varphi}_{s_{k}}$ to maximize $\lambda_k^{(k)}$ when $k<n$ and maximize $\lambda_n^{(k)}$ when $k\geq n$. The greedy algorithm for the sensor placement problem named \emph{minimum nonzero eigenvalue pursuit} (MNEP) is summarized in Algorithm \ref{Algo:MNEP}.

\begin{algorithm}[ht]
\KwIn{$\tilde{\mathbf{\Phi}} = [\boldsymbol\varphi_1, \boldsymbol\varphi_2,..., \boldsymbol\varphi_N]^\mathrm{T}\in \mathbb{R}^{N\times n}$}
\KwOut{$\mathbf{\Phi} \in\mathbb{ R}^{M\times n}$, $\mathcal{S}$, $M$ }

\textbf{Initialization}: $\mathcal{N}=\{1, 2,..., N\}$, $\mathcal{S}=\emptyset$.\\
\textbf{Determine the first $n-1$ sensing locations}: \linebreak
(a) $\mathbf{\Phi}_0=[~]$, $k = 1$. \linebreak
(b) $\hat s_k= \underset{i\in\mathcal N \setminus \mathcal S}{\arg\,\max}~ \lambda_k(\mathbf{\Phi}_{k-1}^\mathrm{T}\mathbf{\Phi}_{k-1}+\boldsymbol{\varphi}_i\boldsymbol{\varphi}_i^\mathrm{T})$. \linebreak
(c) Update: $\mathcal S = \mathcal S\cup\{\hat s_k\},\; \mathbf{\Phi}_k=[\mathbf{\Phi}_{k-1}^\mathrm{T} \;\; \boldsymbol{\varphi}_{\hat s_k}]^\mathrm{T}$.\linebreak
(d) Set $k=k+1$ and repeat step (b-c) until $k=n$.\\
\textbf{Determine the remaining sensing locations}:\linebreak
(a) $\hat s_k= \underset{i\in\mathcal N \setminus \mathcal S}{\arg\,\max}~ \lambda_n(\mathbf{\Phi}_{k-1}^\mathrm{T}\mathbf{\Phi}_{k-1}+\boldsymbol{\varphi}_i\boldsymbol{\varphi}_i^\mathrm{T})$. \linebreak
(b) Update: $\mathcal S = \mathcal S\cup\{\hat s_k\},\; \mathbf{\Phi}_k=[\mathbf{\Phi}_{k-1}^\mathrm{T} \;\; \boldsymbol{\varphi}_{\hat s_k}]^\mathrm{T}$.\linebreak
(c) If $\lambda_n^{(k)}\geq \gamma$ return $\mathcal{S}$, $M=k$, and $\mathbf{\Phi} = \mathbf{\Phi}_k$, else set \hspace*{4mm} $k = k+1$ and repeat step (a-b).
\caption{minimum nonzero eigenvalue pursuit}
\label{Algo:MNEP}
\end{algorithm}

We determine the sensing locations one-by-one. For the first $n-1$ sensors, the $k$-th sensing location can be obtained from the optimization problem in step 2(b). To solve this optimization problem, we traverse all the unselected observation vectors and find the one that maximizes $\lambda_k^{(k)}$. After $n-1$ sensing locations have been determined, we find the remaining sensing locations by solving the optimization problem in step 3(a), which is similar to the previous one but maximizes $\lambda_n^{(k)}$, i.e. the minimum eigenvalue of $\mathbf{\Psi}_k$. Meanwhile, we check the constraint in \eqref{eq:WCEVOptimization} after each sensing location is determined. If the constraint is satisfied, stop the algorithm.

When determining the $k$-th sensing location, we need to traverse $N-k+1$ rows of $\tilde{\mathbf{\Phi}}$, and evaluate the minimum nonzero eigenvalue of $\mathbf{\Psi}_k$ for each case. Solving eigenvalue problems for all $N-k+1$ cases is computationally expensive. Can we find a simpler alternative criterion for the observation vectors in $\{\boldsymbol{\varphi}_{i}\}_{i=1}^N$ to avoid solving eigenvalue problems for all possible dual observation matrices?

In what follows, we provide another criterion, i.e. the magnitude of the projection of $\boldsymbol{\varphi}_{s_k}$ onto the \emph{minimum eigenspace} of $\mathbf{\Psi}_{k-1}$, and it is shown to satisfy the three aforementioned conditions.
Compared with the MNEP, the greedy algorithm via optimizing the new criterion is computationally more efficient and effective.
The definition of minimum eigenspace will be given later.

\subsection{Maximal projection on minimum eigenspace (MPME)} \label{SubSection:MPME}

Sensor placement is about selecting proper observation vectors to guarantee that the eigenvalues of $\mathbf{\Psi}_M$ meet certain requirements. To understand how the observation vectors $\{\boldsymbol{\varphi}_{s_{i}}\}_{i=1}^k$ affect the eigenvalues of $\mathbf{\Psi}_k$, we introduce the following theorem.
\begin{theorem}
\label{Theorem:Eigenvalue&Projection}
For any matrix $\mathbf{\Phi}_k\in\mathbb{R}^{k\times n}$, the symmetric matrix $\mathbf{\Psi}_k = {\mathbf{\Phi}}_k^\mathrm{T}{\mathbf{\Phi}}_k$ has a nonincreasing eigenvalue sequence $\{\lambda_i^{(k)}\}_{i=1}^n$, and \begin{equation}
\lambda_i^{(k)} = \|{\mathbf{\Phi}}_k\mathbf{u}_i^{(k)}\|_2^2 =\sum_{j=1}^k(\boldsymbol{\varphi}_{s_{j}}^{\mathrm{T}}\mathbf{u}_i^{(k)})^2  \label{eq:eigenvalueEQ}
\end{equation}
where $\mathbf{u}_i^{(k)}$ is the normalized eigenvector associated with $\lambda_i^{(k)}$.
\end{theorem}
\begin{IEEEproof}
See Appendix A.
\end{IEEEproof}

In \eqref{eq:eigenvalueEQ}, $\boldsymbol{\varphi}_{s_{j}}^{\mathrm{T}}\mathbf{u}_i^{(k)}$
represents the magnitude of the projection of $\boldsymbol{\varphi}_{s_{j}}$ onto the eigenspace $\mathbf{span}(\mathbf{u}_i^{(k)})$, which is associated with the eigenvalue $\lambda_i^{(k)}$. Theorem 2 shows that the eigenvalue of $\mathbf{\Psi}_k$ equals the square summation of the projections of all columns of $\mathbf{\Phi}_k^\mathrm{T}$ onto its eigenspace.

For any $k\geq n$, the minimum eigenvalue of $\mathbf{\Psi}_k$, i.e. $\lambda_n^{(k)}$, equals the square summation of the projections of  $\{\boldsymbol{\varphi}_{s_{i}}\}_{i=1}^k$ onto the eigenspace associated with the minimum eigenvalue. However, before the $k$-th sensing location is determined, the eigenspace of $\mathbf{\Psi}_k$ is unknown. Therefore, to assess the contribution of the all unselected  observation vectors for $\lambda_n^{(k)}$, we need to solve the eigenvalue problems for all potential cases, like the MNEP, which is computationally expensive.

To analyze the relation between the observation vectors in $\{\boldsymbol{\varphi}_{s_{i}}\}_{i=1}^k$  and the eigenspace associated with the minimum eigenvalue of $\mathbf{\Psi}_k$, we present the following theorem.

\begin{theorem}
\label{Theorem:S_and_N}
The normalized eigenvector associated with $\lambda_n^{(k)}$ of $\mathbf{\Psi}_k$
\begin{equation}
\mathbf{u}_n^{(k)} = \underset{\|\mathbf{x}\|_2=1}{\mathrm{arg}\;\mathrm{min}} \|\mathbf{\Phi}_k\mathbf{x}\|_2^2 \label{eq:EigenvectorDirection}
\end{equation}
and one \emph{sufficient and necessary condition} of $\lambda_n^{(k)}>\gamma$ is that for any nonzero normalized vector $\mathbf{x}\in\mathbb{R}^n$,
\begin{equation}
\|\mathbf{\Phi}_k\mathbf{x}\|_2^2>\gamma \label{eq:S&N}
\end{equation}
\end{theorem}

\begin{IEEEproof}
See Appendix B.
\end{IEEEproof}

For any $k>n$, equation \eqref{eq:EigenvectorDirection} shows that $\mathbf{span}(\mathbf{u}_n^{(k-1)})$ is the subspace onto which the square summation of the projections of $\{\boldsymbol{\varphi}_{s_{i}}\}_{i=1}^{k-1}$ is minimum. However, if the minimum eigenvalue of $\mathbf{\Psi}_{k-1}$, i.e. $\lambda_n^{(k-1)}$, is a multiple eigenvalue with multiplicity $\mu_n$, considering equations \eqref{eq:eigenvalueEQ} and \eqref{eq:EigenvectorDirection} we can find that the optimization problem in \eqref{eq:EigenvectorDirection} has $\mu_n$ different optimal solutions, which are exactly the normalized eigenvectors associated with the minimum eigenvalue of $\mathbf{\Psi}_{k-1}$, i.e. $\lambda_{n-\mu_n+1}^{(k-1)}=\lambda_{n-\mu_n+2}^{(k-1)}=...=\lambda_n^{(k-1)}$.

For any $k\leq n$, the projection of any vector in $\{\boldsymbol{\varphi}_{s_{i}}\}_{i=1}^{k-1}$ onto the null space of $\mathbf{\Phi}_{k-1}$, i.e. $\mathbf{null}(\mathbf{\Phi}_{k-1})$, is zero. It is clear that the square summation of the projections of $\{\boldsymbol{\varphi}_{s_{i}}\}_{i=1}^{k-1}$ onto any other subspace except the subspace of $\mathbf{null}(\mathbf{\Phi}_{k-1})$ is nonzero. Therefore, $\mathbf{null}(\mathbf{\Phi}_{k-1})$ is exactly the subspace with the highest dimension onto which the square summation of the projections of all vectors in $\{\boldsymbol{\varphi}_{s_{i}}\}_{i=1}^{k-1}$ is minimum.

For simplicity, we introduce a new concept, the \emph{minimum eigenspace}, as follows.
\begin{definition}
For any positive semi-define symmetric matrix $\mathbf{A}\in \mathbb{R}^{n\times n}$ with the nonincreasing eigenvalue sequence $\{\lambda_i(\mathbf{A})\}_{i=1}^n$, the \emph{minimum eigenspace} of $\mathbf{A}$ is the eigenspace associated with all the minimum eigenvalues of $\mathbf{A}$, i.e.
\begin{equation*}
\mathbb{U}_{k:n}(\mathbf{A}) = \mathbf{span}(\mathbf{u}_k, \mathbf{u}_{k+1},..., \mathbf{u}_n)
\end{equation*}
where $\mathbf{u}_i$ is the eigenvector associated with $\lambda_i(\mathbf{A})$, and $\lambda_{k-1}(\mathbf{A})>\lambda_{k}(\mathbf{A})=\lambda_n(\mathbf{A})$.
\end{definition}

Equation \eqref{eq:S&N} implies that to meet the requirement on $\lambda_n(=\lambda_n^{(M)})$, i.e. $\lambda_n \geq \gamma$, the square summation of the projections of $\{\boldsymbol{\varphi}_{s_{k}}\}_{k=1}^M$ onto any non-trivial subspace of $\mathbb{R}^n$ should be larger than $\gamma$.
Therefore, it is reasonable that when determining the  $k$-th sensing location we select the observation vector ${\boldsymbol{\varphi}}_{s_{k}^*}$ that has the largest projection onto the non-trivial subspace onto which the square summation of the projections of  $\{\boldsymbol{\varphi}_{s_{i}}\}_{i=1}^{k-1}$ is minimum.

Therefore, if $k\leq n$  we can select the $k$-th sensing location whose observation vector has the largest projection onto the null space of $\mathbf{\Phi}_{k-1}$. It is easily found that the null space of $\mathbf{\Phi}_{k-1}$ is exactly the minimum eigenspace of $\mathbf{\Psi}_{k-1}$, i.e.  $\mathbf{null}(\mathbf{\Phi}_{k-1})=\mathbb{U}_{k:n}(\mathbf{\Psi}_{k-1})$. If $k>n$, the square summation of the projections of $\{\boldsymbol{\varphi}_{s_{i}}\}_{i=1}^{k-1}$ onto the $\mu_n$ subspaces  $\mathbf{span}(\mathbf{u}_{n-\mu_n+1}^{(k-1)})$,  $\mathbf{span}(\mathbf{u}_{n-\mu_n+2}^{(k-1)})$,...,  and $\mathbf{span}(\mathbf{u}_n^{(k-1)})$ are equal and minimum. Generally, $\lambda_n^{k-1}$ is a simple eigenvalue and $\mu_n=1$. If $\mu_n>1$   we can select the $k$-th observation vector that has the largest projection onto the spanned subspace $\mathbf{span}(\mathbf{u}_{n-\mu_n+1}^{(k-1)}, \mathbf{u}_{n-\mu_n+2}^{(k-1)},...,\mathbf{u}_n^{(k-1)})$, which is exactly the minimum eigenspace of $\mathbf{\Psi}_{k-1}$.

In summary, for all $k\geq 1$, we can determine the $k$-th sensing location by maximizing the projection of the observation vector on the minimum eigenspace of the current dual observation matrix $\mathbf{\Psi}_{k-1}$.

For simplicity, we denote
\begin{equation}
\mathbf{z}=(\mathbf{U}^{(k-1)})^\mathrm{T}\boldsymbol{\varphi}_{s_{k}}
\label{eq:zdifinition}
\end{equation}
where $\mathbf{U}^{(k-1)}=[\mathbf{u}_{1}^{(k-1)}, \mathbf{u}_{2}^{(k-1)},...,\mathbf{u}_{n}^{(k-1)}]$, and let $z_i$ be the $i$-th component of $\mathbf{z}$. It is clear that the square of the projection of $\boldsymbol{\varphi}_{s_{k}}$ onto the minimum eigenspace of $\mathbf{\Psi}_{k-1}$ is
\begin{equation*}
\zeta_k = \sum_{i=n-\mu_n+1}^{n}z_i^2=\left\{
\begin{aligned}
    \sum_{i=k}^{n} z_{i}^2 \;\;\quad \quad&\textmd{if}&  k<n\\
    \sum_{i=n-\mu_n+1}^{n}z_i^2\quad &\textmd{if}&  k\geq n
\end{aligned}
\right.
\end{equation*}

Let $\mathbf{P}_{k-1}$ be a projection matrix which can project any vectors in $\mathbb{R}^n$ onto the minimum eigenspace of $\mathbf{\Psi}_{k-1}$. When $k<n$, the minimum eigenspace of $\mathbf{\Psi}_{k-1}$ is the null space of $\mathbf{\Phi}_{k-1}$. Then, we can find that

\begin{equation*}
\mathbf{P}_{k-1}=\mathbf{I}_{n\times n}-\mathbf{R}_{k-1}\mathbf{R}_{k-1}^\mathrm{T}
\end{equation*}
where $\mathbf{R}_{k-1}=\mathbf{orth}(\mathbf{\Phi}^\mathrm{T}_{k-1})$ whose column vectors are obtained from the Gram-Schmidt Orthonormalization of all the column vectors of $\mathbf{\Phi}^\mathrm{T}_{k-1}$, i.e. the vector group $\{\boldsymbol{\varphi}_{s_i}\}_{i=1}^{k-1}$. When $k\geq n$, it is clear that
\begin{equation*}
\mathbf{P}_{k-1}=\mathbf{U}_n^{(k-1)}(\mathbf{U}_n^{(k-1)})^\mathrm{T}
\end{equation*}
where $\mathbf{U}_n^{(k-1)}=[\mathbf{u}_{n-\mu_n+1}^{(k-1)}, \mathbf{u}_{n-\mu_n+2}^{(k-1)},...,\mathbf{u}_{n}^{(k-1)}]$. Then we can obtain that
\begin{equation*}
\zeta_k = \|\mathbf{P}_{k-1}\boldsymbol{\varphi}_{s_k}\|_2^2
\end{equation*}

Apparently $\zeta_k$ can be obtained if $\mathbf{\Phi}_{k-1}$ and $\boldsymbol{\varphi}_{s_{k}}$ are known, i.e. $\zeta_k$ satisfies the first condition of the new criterion used to replace $\lambda_n$. If $\zeta_k\neq 0$ for any $k<n$, the projection of $\boldsymbol{\varphi}_{s_k}$ onto $\mathbf{null}(\mathbf{\Phi}_{k-1})$ is nonzero; therefore, $\boldsymbol{\varphi}_{s_k}$ is independent with all the vectors in $\{\boldsymbol{\varphi}_{s_i}\}_{i=1}^{k-1}$, which implies that $\zeta_k$ satisfies the second condition. Next, we utilize the following theorem to show that $\zeta_k$ satisfies the third condition.

\begin{theorem} \label{Theorem:PCorrelated}
Given any observation matrix $\mathbf{\Phi}_k= [\boldsymbol{\varphi}_{s_1}, \boldsymbol{\varphi}_{s_2},$ $..., \boldsymbol{\varphi}_{s_k}]^\mathrm{T} \in \mathbb{R}^{k\times n}$, and its corresponding dual observation matrix  $\mathbf{\Psi}_k = {\mathbf{\Phi}}_k^\mathrm{T}{\mathbf{\Phi}}_k$ with a nonincreasing eigenvalue sequence $\{\lambda_i^{(k)}\}_{i=1}^n$.

If $k\leq n$ and $\mathbf{\Phi}_k$ is full row-rank, then
\begin{subequations}
\begin{align}
\lambda_{k}^{(k)}&=\frac{\zeta_k}{1+\sum\limits_{i=1,z_i\neq 0}^{k-1}\frac{z_i^2}{\lambda_i^{(k-1)}-\lambda_{k}^{(k)}}} \label{eq:MinmultiEigenvalue}\\
\lambda_{k+1}^{(k)}&= \lambda_{k+2}^{(k)}...=\lambda_{n}^{(k)}=0 \label{eq:multi-Eigenvalue}
\end{align}
\end{subequations}
and $\lambda_k^{(k)}$ is monotonically increasing w.r.t. $\lambda_{k-1}^{(k-1)}$.

If $k\geq n$, then
\begin{subequations}
\begin{align}
\lambda_{n-\mu_n+1}^{(k)}&=\lambda_{n-\mu_n+1}^{(k-1)}+\frac{\zeta_k} {1+\sum\limits_{i=1,z_i\neq0}^{n-\mu_n}\frac{z_i^2}{\lambda_i^{(k-1)}-\lambda_{n-\mu_n+1}^{(k)}}}\label{eq:MinEigenvalue}\\
\lambda_{n-\mu_n+2}^{(k)}&= \lambda_{n-\mu_n+3}^{(k)}...=\lambda_{n}^{(k)}=\lambda_n^{(k-1)} \label{eq:multi-Eigenvalue_n}
\end{align}
\end{subequations}
and for any $M\geq n$, $\lambda_n^{(M)}$ is monotonically increasing w.r.t. $\zeta_k$ for all $k\leq M$.
\end{theorem}
\begin{IEEEproof}
See Appendix C.
\end{IEEEproof}

This theorem shows that $\lambda_n$ is monotonically increasing w.r.t. $\zeta_k$. Accordingly, WCEV is monotonically decreasing w.r.t. $\zeta_k$. It is clear that $\zeta_k$ can be another choice of the criterion instead of $\lambda_n$ to optimize the $k$-th sensing location. Therefore, to find the $k$-th sensing location, we maximize the magnitude of the projection of $\boldsymbol{\varphi}_{s_k}$ onto the minimum eigenspace of $\mathbf{\Psi}_{k-1}$ instead of maximizing the minimum nonzero eigenvalue of $\mathbf{\Psi}_k$. The greedy sensor placement algorithm, which we call \emph{maximal projection on minimum eigenspace} (MPME), is given in Algorithm \ref{algo2}.

\begin{algorithm}[ht]
\KwIn{$\tilde{\mathbf{\Phi}} = [\boldsymbol\varphi_1, \boldsymbol\varphi_2,..., \boldsymbol\varphi_N]^\mathrm{T}\in \mathbb{R}^{N\times n}$}
\KwOut{$\mathbf{\Phi} \in\mathbb{ R}^{M\times n}$, $\mathcal{S}$ , $M$}

\textbf{Initialization}: $\mathcal{N}=\{1, 2,..., N\}$, $\mathcal{S}=\emptyset$. \\

\textbf{Determine the first $n-1$ sensing locations}: \linebreak
(a) Set $\mathbf{\Phi}_0=[~]$, $\mathbf{P}_0=\mathbf{I}_{n\times n}$ and $k=1$. \linebreak
(b) $\hat s_k = \underset{i\in\mathcal N \setminus \mathcal S}{\arg\,\max}~\|\mathbf{P}_{k-1}\boldsymbol{\varphi}_i\|_2^2$. \linebreak
(c) Update: $\mathcal S = \mathcal S\cup\{\hat s_k\},\; \mathbf{\Phi}_k=[\mathbf{\Phi}_{k-1}^\mathrm{T} \;\; \varphi_{\hat s_k}]^\mathrm{T}$, \linebreak
\hspace*{17mm}$\mathbf{R}_k =\mathrm{\mathbf{orth}}(\mathbf{\Phi}_k^\mathrm{T})$, $\mathbf{P}_k=\mathbf{I}_{n\times n}-\mathbf{R}_{k}\mathbf{R}_{k}^\mathrm{T}$. \linebreak
(d) Set $k=k+1$ and repeat step (b-c) until $k=n$.\\

\textbf{Determine the remaining sensing locations}:\linebreak
(a) $\hat s_k = \underset{i\in\mathcal N \setminus \mathcal S}{\arg\,\max}~\|\mathbf{P}_{k-1}\varphi_i\|_2^2$. \linebreak
(b) Update: $\mathcal S = \mathcal S\cup\{\hat s_k\},\; \mathbf{\Phi}_k=[\mathbf{\Phi}_{k-1}^\mathrm{T} \;\; \varphi_{\hat s_k}]^\mathrm{T}$,
\hspace*{17 mm}$\mathbf{ \Phi}_k^\mathrm{T}\mathbf{\Phi }_k= \mathbf{U}\mathbf{\Lambda}_k\mathbf{U}^\mathrm{T}$, $\mathbf{P}_{k} = \mathbf{U}_n\mathbf{U}_n^\mathrm{T}$.\linebreak
(c) If $\lambda_n^{(k)}\geq \gamma$ return $\mathcal{S}$, $M=k$, and $\mathbf{\Phi} = \mathbf{\Phi}_k$, else set \hspace*{4 mm} $k = k+1$ and repeat step (a-b).
\caption{maximal projection on minimum eigenspace}
\label{algo2}
\end{algorithm}

The $k$-th sensing location is obtained from the optimization problem in step 2(b) or step 3(a). To solve the optimization problem, we traverse all the unselected observation vectors and find the one that maximizes $\zeta_k$. Meanwhile, if $k\geq n$, we check the constraint in \eqref{eq:WCEVOptimization} after each sensing location is determined. If the constraint is satisfied, stop the algorithm.

\begin{figure*}[htb]
    \centering
    \begin{subfigure}[b]{0.5\textwidth}
        \centering
        \includegraphics[width=\textwidth]{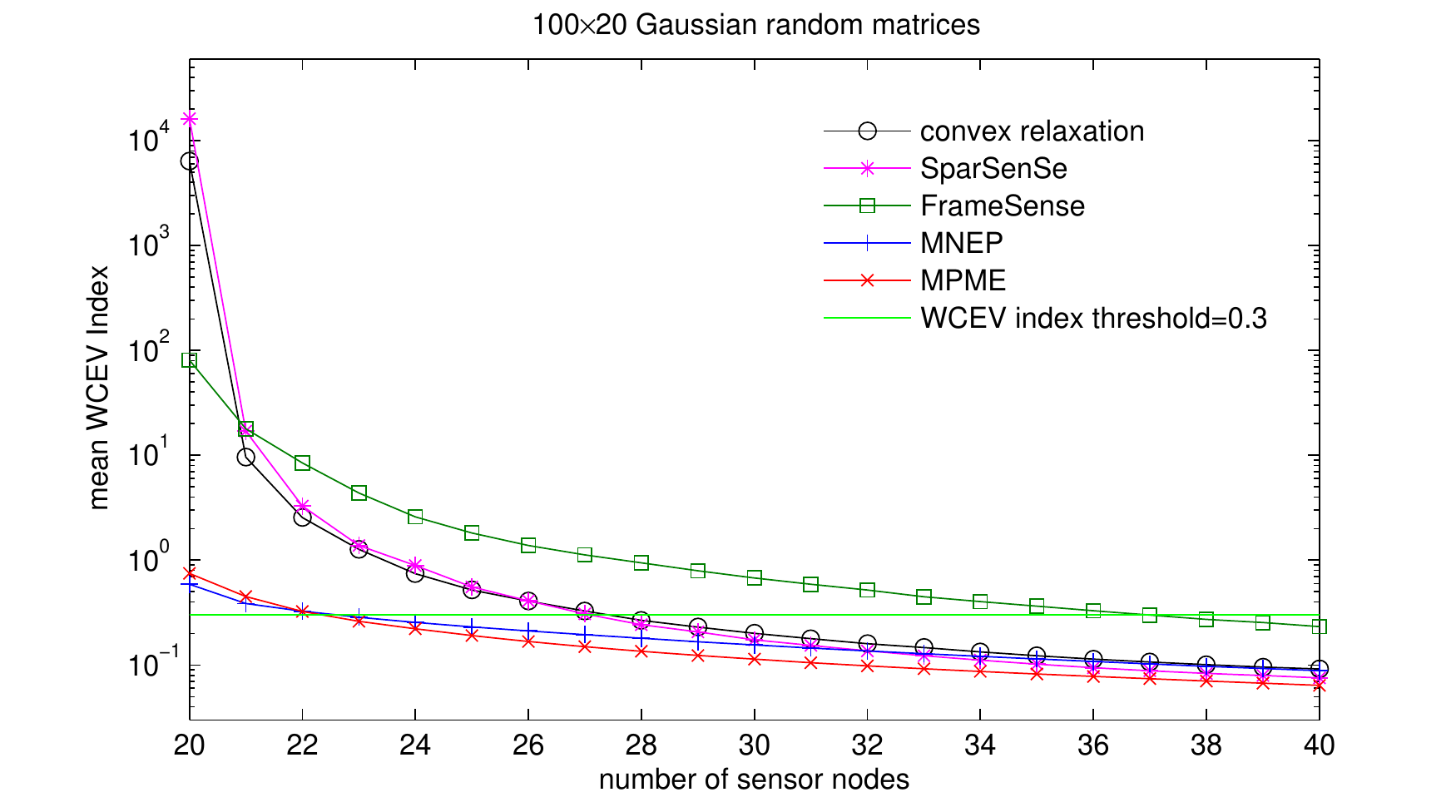}
    \end{subfigure}%
    \begin{subfigure}[b]{0.5\textwidth}
        \centering
        \includegraphics[width=\textwidth]{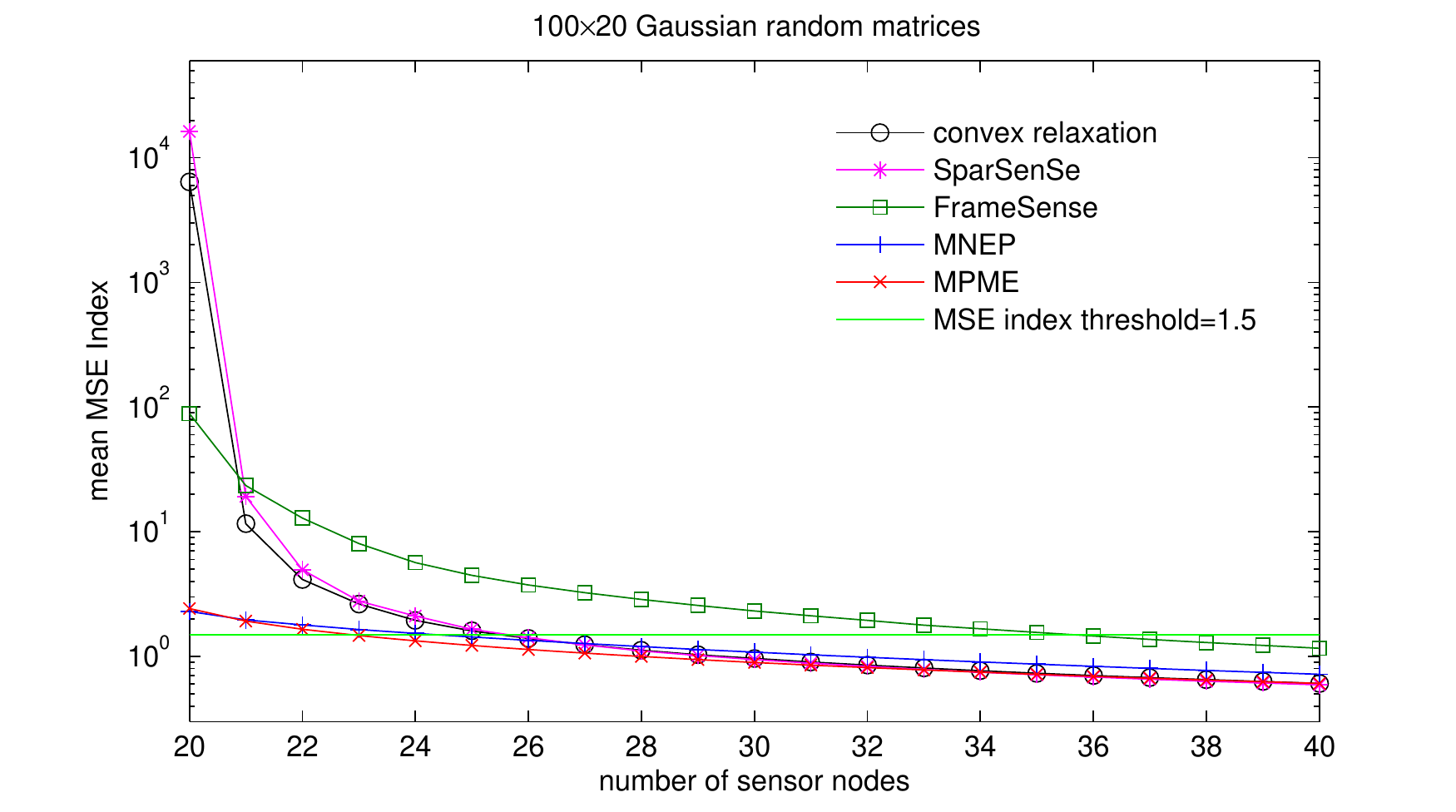}
    \end{subfigure}
    \begin{subfigure}[b]{0.5\textwidth}
    \centering
    \includegraphics[width=1\textwidth]{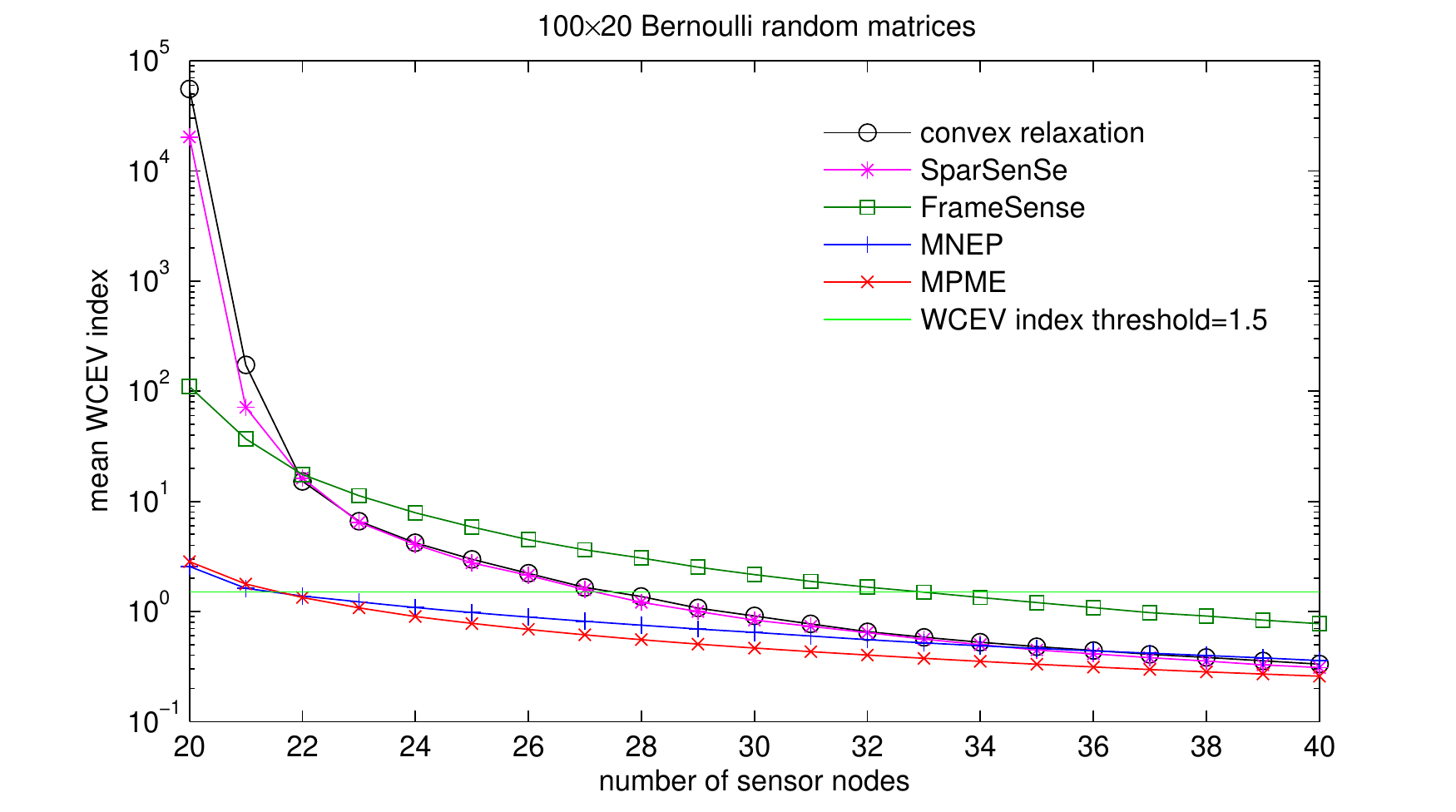}
    \end{subfigure}%
    \begin{subfigure}[b]{0.5\textwidth}
        \centering
        \includegraphics[width=\textwidth]{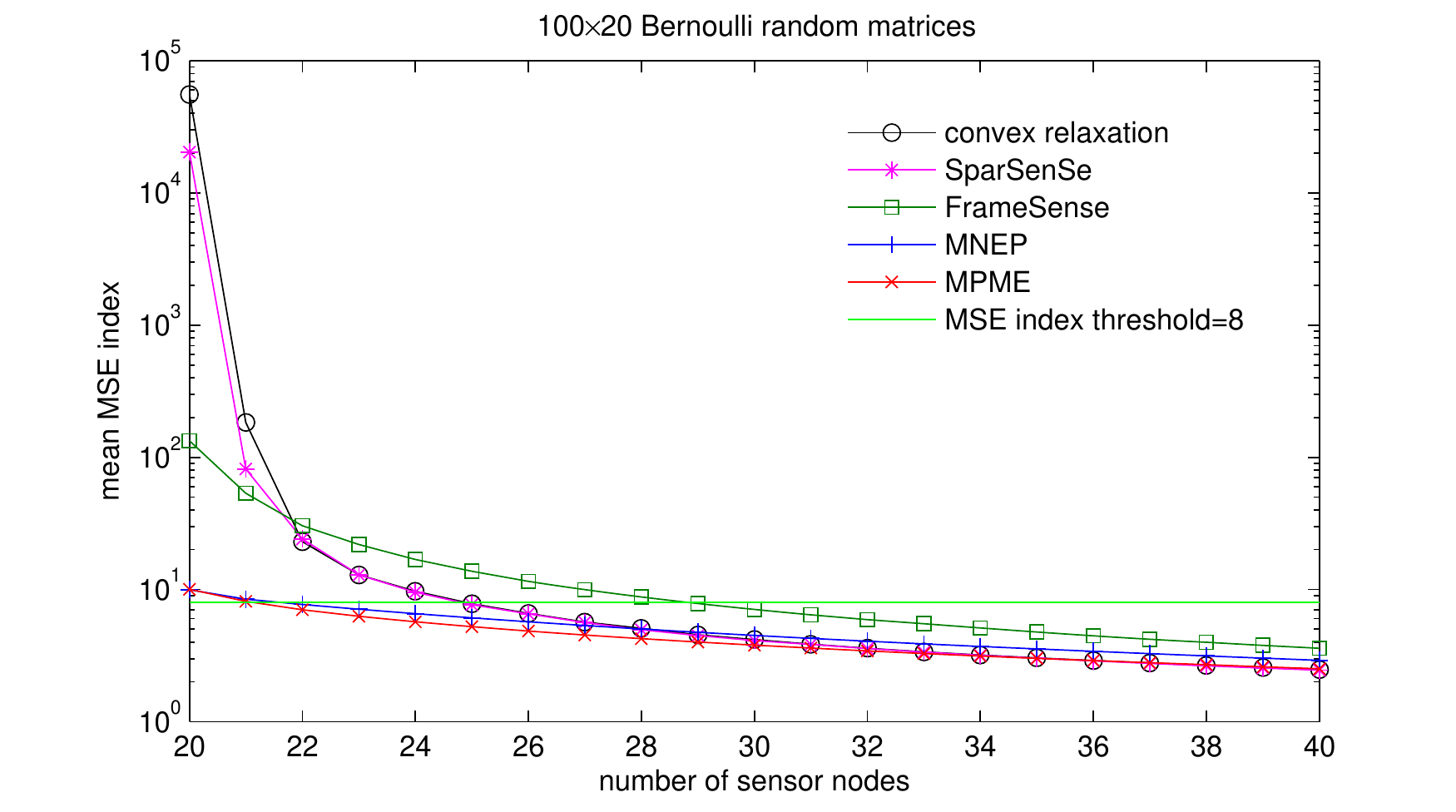}
    \end{subfigure}
    \caption{The performance comparison between the MPME and other sensor placement methods for \emph{Example} \ref{example:generalGaussion} and \emph{Example} \ref{example:generalBernoulli}. For the MPME and the MNEP we do not consider the stopping criteria but show the mean WCEV index and the mean MSE index w.r.t. the number of sensor nodes, increasing from 20 to 40. For the SparSenSe method, we do not consider the minimum number of required sensor nodes but determine the sensing locations like the convex relaxation method by finding the indices of the $M$ largest elements of $\mathbf{w}^*$. }
    \label{Fig:Performance}
\end{figure*}

\subsection{Discussions about MNEP \& MPME}\label{SubSection:discussion}

In both algorithms we need to solve the optimization problem in step 2(b) or step 3(a). In Algorithm \ref{Algo:MNEP}, to solve the optimization problem, we need to evaluate the minimum nonzero eigenvalues of all $N-k+1$ possible dual observation matrices, while in Algorithm \ref{algo2} we compute the projection of $N-K+1$ possible observation vectors onto the minimum eigenspace of $\mathbf{\Psi}_{k-1}$. The computational cost to find the eigenvalue of an $n\times n$ matrix is much more expensive than projecting one vector in $\mathbb{R}^n$ onto a known subspace. Therefore, the MPME is computationally much more efficient.

Equations \eqref{eq:MinmultiEigenvalue} shows that $\lambda_k^{(k)}$ is monotonic increasing w.r.t. $\zeta_k$ and decreasing w.r.t. $z_i^2$ for all $i<k\leq n$. Generally $\lambda_n^{(k-1)}$ is a simple eigenvalue when $k>n$, i.e. $\mu_n=1$, and \eqref{eq:MinEigenvalue}-\eqref{eq:multi-Eigenvalue_n} can be simplified as
\begin{equation}
\lambda_{n}^{(k)}=\lambda_{n}^{(k-1)}+\frac{\zeta_k} {1+\sum\limits_{i=1,z_i\neq0}^{n-1}\frac{z_i^2}{\lambda_i^{(k-1)}-\lambda_{n}^{(k)}}}\label{eq:MinSimpleEigenvalue}
\end{equation}
Equation \eqref{eq:MinSimpleEigenvalue} shows that $\lambda_n^{(k)}$ is monotonic increasing w.r.t. $\zeta_k$ and decreasing w.r.t. $z_i^2$ for all $i<n\leq k$. Therefore, the MNEP algorithm prefers to select the observation vector with large $\zeta_k$ and small $z_i^2$, which may not be with the largest $\zeta_k$ but achieves a balance between a large $\zeta_k$ and a small $z_i^2$. It is clear that $\sum_{i=1}^n z_i^2 =\|\boldsymbol{\varphi}_{s_k}\|_2^2$ and $\sum_{i=1}^k\|\boldsymbol{\varphi}_{s_i}\|_2^2=\sum_{i=1}^n \lambda_i^{(k)}$; therefore, small $z_i^2$ means small increment of $\lambda_i^{(k)}$ from $\lambda_i^{(k-1)}$ for all $i<k$ when $k<n$, and all $i<n$ when $k\geq n$. Additionally, \eqref{eq:MinmultiEigenvalue} and \eqref{eq:MinSimpleEigenvalue} show that both $\lambda_k^{(k)}$ and $\lambda_n^{(k)}$ are monotonically increasing w.r.t. $\lambda_i^{(k-1)}$. Therefore, selecting the observation vector with small $z_i^2$ leads to a relatively smaller $\lambda_{k_+}^{(k_+)}$ and $\lambda_n^{(k_+)}$, where $k_+>k$, and the MPME probably outperforms the MNEP in terms of finding the largest $\lambda_n^{(k)}$ for $k\geq n$.

In MPME we maximize $\zeta_k$, which can guarantee a large minimum nonzero eigenvalue of the updated dual observation matrix. In MNEP we directly maximize the minimum nonzero eigenvalue of the current dual observation matrix. Therefore, both algorithms guarantee that $\lambda_n^{(k)}$ is not much smaller than $\lambda_i^{(k)}$ for $i<n$ and $k\gtrapprox n$, which implies that both algorithms can guarantee that $\mathbf{\Psi}_k$ is well-conditioned when $k\gtrapprox n$. Here, $k\gtrapprox n$ means $k$ is slightly larger than $n$.

In both algorithms, the minimum number of required sensor nodes is determined by judging whether $\lambda_n^{(k)}\geq\gamma$ is satisfied after each sensing location is determined when $k\geq n$. It is clear that in both algorithms the constraint in \eqref{eq:WCEVOptimization} is only considered in step 3(c), i.e. the last sub-step. In other words, the constraint in \eqref{eq:WCEVOptimization} is only used to judge whether the number of sensor nodes is enough.

\begin{remark}
We can change the constraint in \eqref{eq:WCEVOptimization} to other constraints described by MSE or $\mathbf{{det}}(\mathbf{\Sigma})$, if other measures are more interesting. Correspondingly, in step 3(c) we check the new constraint.
\end{remark}

\section{effectiveness of the MPME algorithm}

In this section, we provide four examples to illustrate the effectiveness of the MPME. The detailed analysis of the comparisons with the state-of-the-art, i.e. the convex relaxation \cite{joshi2009sensor}, SparSenSe \cite{jamali2014sparsity}, and FrameSense \cite{ranieri2014near} will also be presented.

\begin{example} \label{example:generalGaussion}
$\tilde{\mathbf{\Phi}} \in \mathbb{R}^{100\times 20}$ is a Gaussian random matrix with independent components $\varphi_{ij}\sim \mathscr{N}(0,1)$, and the variance of the sensor noise $\sigma^2 =1$.
\end{example}

\begin{example} \label{example:generalBernoulli}
$\tilde{\mathbf{\Phi}} \in \mathbb{R}^{100\times 20}$ is a Bernoulli random matrix with independent components $\varphi_{ij}\sim \mathscr{B}(1,0.5)$ with $\mathscr{B}$ representing the Binomial distribution, and the variance of the sensor noise $\sigma^2 =1$.
\end{example}

The mean WCEV index and the mean MSE index of 200 Monte-Carlo simulation run results for the two examples are given in Fig. \ref{Fig:Performance}. The optimization problems \eqref{eq:convexRelaxation} and \eqref{eq:SparSenSe} are solved using the SDPT3 solver \cite{toh1999sdpt3} and CVX toolbox \cite{Boydcvx}, respectively. Actually, the SDPT3 solver is used as the computational engine of the CVX toolbox.

For the $i$-th ($1\leq i\leq 200$) simulation run, we determine the observation matrix $\mathbf{\Phi}_k^{(i)}$ ($20\leq k\leq 40$) based on the random signal representation matrix $\tilde{\mathbf{\Phi}}^{(i)}$, and obtain the following MSE index and WCEV index from \eqref{eq:MSE} and \eqref{eq:WCEV}, respectively:
\begin{eqnarray}
\mathrm{MSE}_k^{(i)} = \mathbf{tr}((\mathbf{\Psi}_k^{(i)})^{-1}) \label{eq:MSEindex}\\
\mathrm{WCEV}_k^{(i)}= \lambda_{\mathrm{max}}((\mathbf{\Psi}_k^{(i)})^{-1}) \label{eq:WCEVindex}
\end{eqnarray}
where $\mathbf{\Psi}_k^{(i)}=(\mathbf{\Phi}_k^{(i)})^\mathrm{T}\mathbf{\Phi}_k^{(i)}$. Then, the mean MSE index and the mean WCEV index of 200 Monte-Carlo simulation run results are given by
\begin{eqnarray*}
\overline{\mathrm{MSE}}_k = \frac{1}{200}\sum\nolimits_{i=1}^{200} \mathrm{MSE}_k^{(i)} \\
\overline{\mathrm{WCEV}}_k= \frac{1}{200}\sum\nolimits_{i=1}^{200} \mathrm{WCEV}_k^{(i)}
\end{eqnarray*}

It is shown in Fig. \ref{Fig:Performance} that for both examples, the MPME outperforms all the other methods in terms of the mean WCEV index and the mean MSE index. If the same number of sensor nodes are used, the MPME can provide the best results of linear inverse problems as compared with the other methods.

If we set the WCEV index threshold to be 0.3 (i.e. $\gamma = 10/3$) for \emph{Example} \ref{example:generalGaussion}, the top left figure shows that the minimum number of required sensor nodes $M_{\mathrm{MNEP}}=M_{\mathrm{MPME}}=23$, $M_{\mathrm{convex\_relaxation}}=M_{\mathrm{SparSenSe}}=28$, and $M_{\mathrm{FrameSense}}=37$. If we set the MSE index threshold to be 1.5, the top right figure shows that the minimum number of required sensor nodes  $M_{\mathrm{MPME}}=23$, $M_{\mathrm{MNEP}}=25$, $M_{\mathrm{convex\_relaxation}}=M_{\mathrm{SparSenSe}}=26$, and  $M_{\mathrm{FrameSense}}=36$. Therefore, to meet the accuracy requirement, the proposed MPME algorithm requires the least number of sensor nodes. For the Bernoulli random data matrix we can easily obtain the same conclusion from the bottom two figures.

Fig. \ref{Fig:Performance}  shows that MPME outperforms MNEP, which has been analyzed in Section \ref{SubSection:discussion}. Additionally, we find that for all the five algorithms, the improvement of WCEV and MSE  are increasingly insignificant with an increase in the number of sensor nodes. It means that the influence of additional sensor observation declines and its location is not so critical as the previously determined sensing locations. It is clearly shown in Fig. \ref{Fig:Performance} that MPME is much better than the convex relaxation method, SparSenSe, and FrameSense when the number of sensor nodes is very limited, i.e. the MPME method significantly outperforms the state-of-the-art in finding the critical sensing locations, which is analyzed as follows.

\subsection{Comparison with convex relaxation}

Fig. \ref{Fig:Performance} shows that the mean MSE index and the mean WCEV index of the convex relaxation method are much larger than those of the MPME method when the number of sensor nodes is slightly larger than the dimension of the vector to be estimated, i.e. when $k\gtrapprox n = 20$ in this example. Comparing the left two figures with the right two figures in Fig. \ref{Fig:Performance}, we find that when using the convex relaxation method, the WCEV index (i.e. the maximum eigenvalue of ${\mathbf{\Psi}}_k^{-1}$, see equation \eqref{eq:WCEVindex})  contributes the main part of the MSE index (i.e. the trace of ${\mathbf{\Psi}}_k^{-1}$, see equation \eqref{eq:MSEindex}), especially when the number of sensor nodes is small. It means that the maximum eigenvalue of $\mathbf{\Psi}_k^{-1}$ is overwhelmingly larger than the others when $k\gtrapprox n$. In other words, the minimum eigenvalue of $\mathbf{\Psi}_k$ is much smaller than the other eigenvalues, which implies that $\mathbf{\Psi}_k$ is ill-conditioned, and hence the estimated vector $\hat{\boldsymbol\alpha}$ is inaccurate.

\begin{figure}[ht]
\centering
\includegraphics[width=0.5\textwidth]{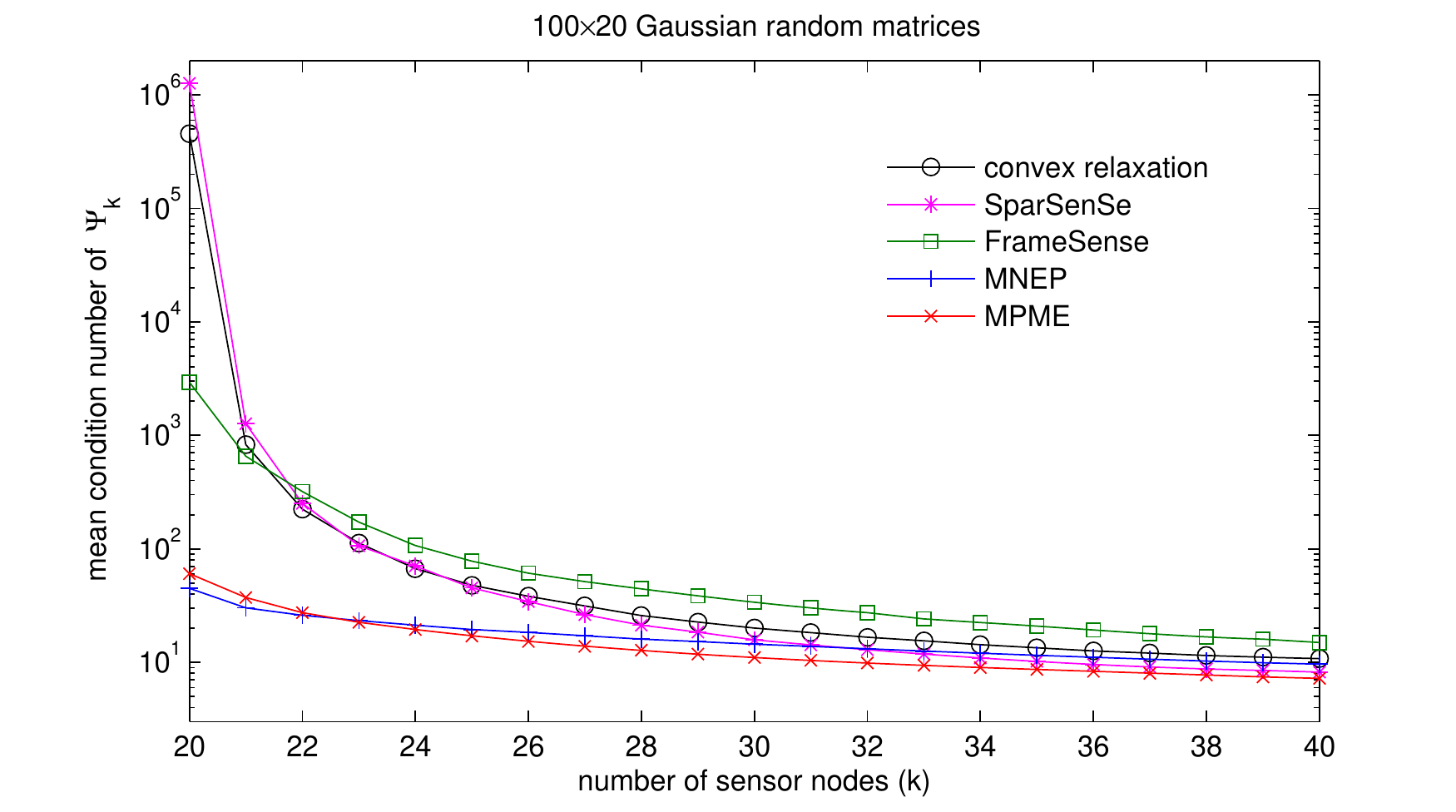}
\caption{The mean condition number of the dual observation matrix $\mathbf{\Psi}_k =\mathbf{\Phi}^\mathrm{T}_k\mathbf{\Phi}_k$ for \emph{Example} \ref{example:generalGaussion}. Here $\mathbf{\Phi}_k$ is obtained from the five sensor placement algorithms.}
\label{Fig:ConditionNO}
\end{figure}

The mean condition number of Monte-Carlo simulation result for \emph{Example} \ref{example:generalGaussion} is shown in Fig. \ref{Fig:ConditionNO}.
For $i$-th ($1\leq i\leq 200$) simulation run, the condition number of $\mathbf{\Psi}_k^{(i)}$ is denoted by $\kappa( \mathbf{\Psi}_k^{(i)})$. The mean condition number is given by
\begin{equation*}
\overline{\kappa}_k = \frac{1}{200}\sum\nolimits_{i=1}^{200} \kappa( \mathbf{\Psi}_k^{(i)})
\end{equation*}
Fig. \ref{Fig:ConditionNO} shows that for $k\leq 23$, $\mathbf{\Psi}_k$ obtained from the convex relaxation method is ill-conditioned, whereas $\mathbf{\Psi}_k$ obtained from  MNEP or MPME is well-conditioned. Accordingly, as shown in Fig. \ref{Fig:Performance}, the mean MSE index and the mean WCEV index of the convex relaxation method are much larger than those of MPME, respectively.

%It is clear that the convex relaxation method is not suitable for the case where the available sensor nodes are very limited.
In practice, the solution of the convex optimization problem \eqref{eq:convexRelaxation}, $\mathbf{w}^*\in [0, 1]^N$, is mapped into $\{0, 1\}^N$ to find the sensing locations. The largest $M$ elements of $\mathbf{w}^*$ are mapped to 1 and other elements are mapped to 0. In such a mapping, the singularity of $\mathbf{\Psi}$ is not considered. The number of sensor nodes is nearer the dimension of the estimated vector $\hat{\boldsymbol\alpha}$, giving a higher probability of $\mathbf{\Psi}$ being ill-conditioned. However, the proposed MNEP and MPME algorithms can guarantee a large minimum nonzero eigenvalue of $\mathbf{\Psi}_k$ and accordingly a well-conditioned $\mathbf{\Psi}_M$ for $M\geq n$.

\subsection{Comparison with SparSenSe}

It is claimed that SparSenSe can determine the minimum number of required sensor nodes by utilizing the sparsity of the variable $\mathbf{w}$ in \eqref{eq:SparSenSe}. With a predefined threshold $\tau$ and the solution of optimization problem \eqref{eq:SparSenSe}, i.e. $\mathbf{w}^*$, if $w_i^*<\tau$ set $w_i^*=0$. The sensing indices then exactly correspond to the nonzero entries of $\mathbf{w}^*$ and the number of nonzero entries is the minimum number of required sensor nodes. This strategy works well for the example in \cite{jamali2014sparsity} in which $\tilde{\mathbf{\Phi}}\in \mathbb{R}^{50\times 2}$, and the largest 3 entries of $\mathbf{w}^*$ are much larger than other elements. We can set $\tau$ as a small value and easily find the largest 3 entries corresponding to the selected sensing indices.

However, this strategy is ineffective if the dimension of the estimated vector $\boldsymbol\alpha$ is large. In the pervious two examples, $\mathbf{w}\in\mathbb{R}^{100}$ with at least 20 nonzero entries is not sparse. We introduce another example to illustrate this problem.

\begin{example} \label{example:generalGaussion1500}
$\tilde{\mathbf{\Phi}} \in \mathbb{R}^{1500\times 20}$ is a Gaussian random matrix with independent components $\varphi_{ij}\sim \mathscr{N}(0,1)$. The variance of the sensor noise $\sigma^2 =1$, and the maximum acceptable MSE index $\gamma'=1.5$.
\end{example}

\begin{figure}[ht]
\centering
\includegraphics[width=0.48\textwidth]{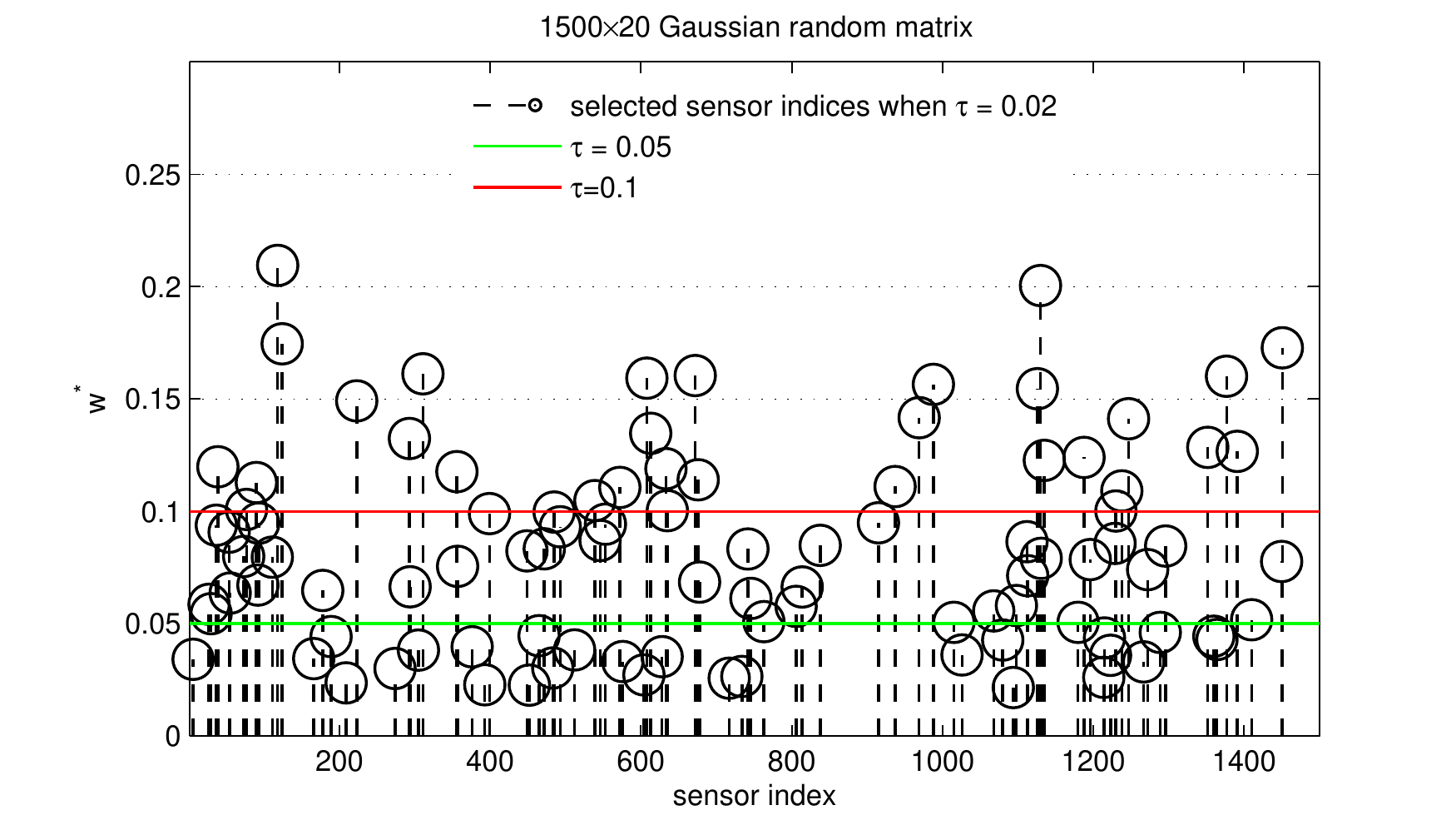}
\caption{Selected sensing indices with SparSenSe for \emph{Example} 3.}
\label{Fig:SensorIndex}
\end{figure}

\begin{figure}[ht]
\centering
\includegraphics[width=0.48\textwidth]{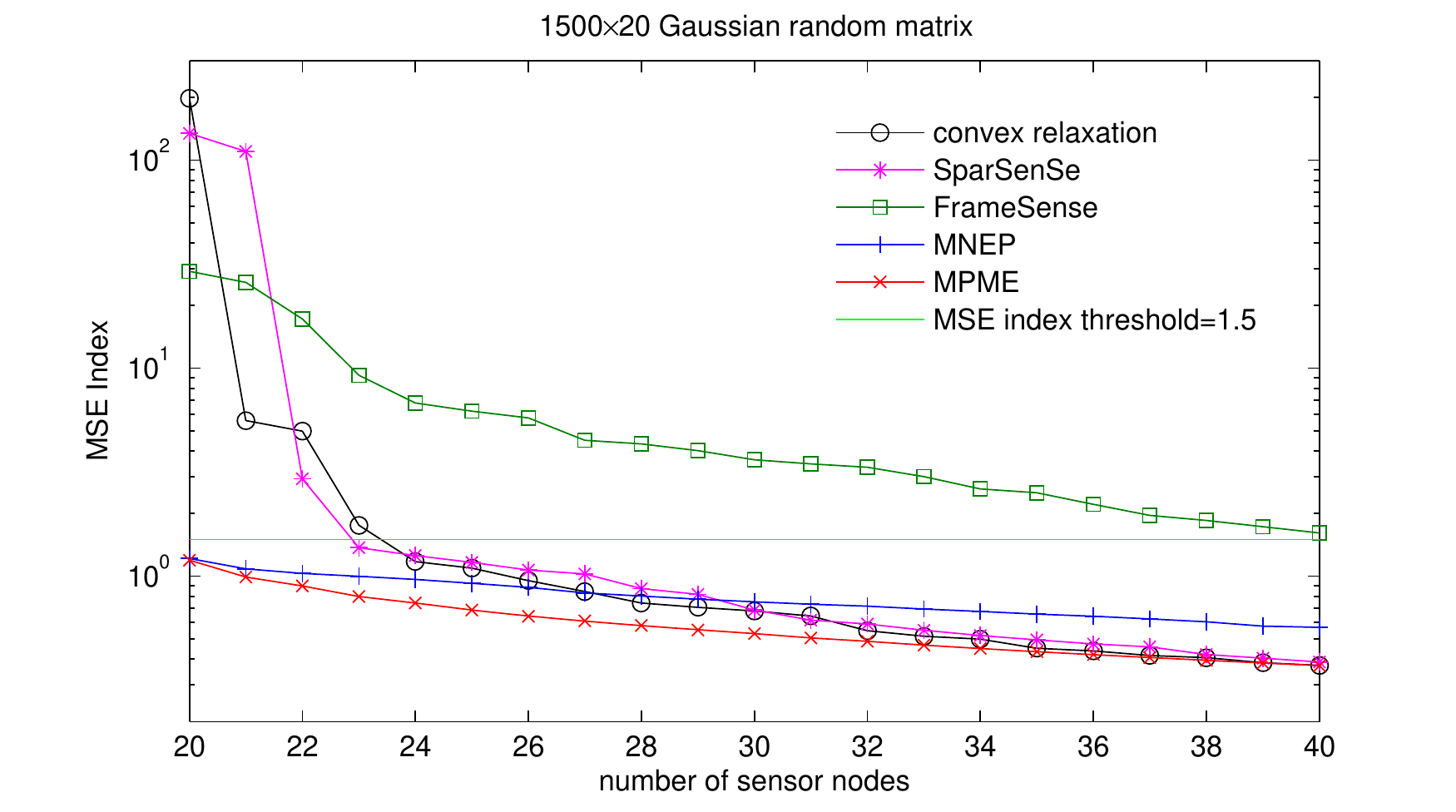}
\caption{The MSE index of $\hat{\boldsymbol\alpha}$ estimated from 20 to 40 sensor observations for \emph{Example} 3.}
\label{Fig:MSE1500}
\end{figure}

In \emph{Example} \ref{example:generalGaussion1500}, $N\gg n$ guarantees that the decision variable $\mathbf{w}$ is sparse. Fig. \ref{Fig:SensorIndex} shows that if $\tau=0.02$, 98 sensor nodes are selected. If $\tau$ is set as 0.05 and 0.1, then 71 and 31 sensor nodes are selected, respectively. We can see the result of SparSenSe from Fig. \ref{Fig:MSE1500} that 31 sensor nodes can guarantee that the MSE index is less than the maximum acceptable MSE index. However, the minimum number of required sensor nodes is 23. For this example, if we set $\tau=\tau^*\in[0.1129, 0.1140]$, the 23 largest elements of $\mathbf{w}^*$ will be selected. In practice, however, the optimal threshold $\tau^*$ is \emph{a prior} unknown; therefore, utilizing the sparsity of $\mathbf{w}$ to determine the minimum number of sensor nodes is ineffective.

In practice, we can set $M=n$ and select the sensing indices that correspond to the $M$ largest entries of $\mathbf{w}^*$. Then, we check whether the accuracy (i.e. WCEV index or MSE index) is acceptable. If not, increase $M$, reselect the sensing indices and recheck the accuracy until the accuracy is acceptable. Using this strategy for \emph{Example} \ref{example:generalGaussion1500}, we can easily find the minimum number of required sensor nodes $M=23$.

Like the convex relaxation method, Fig. \ref{Fig:Performance} and Fig. \ref{Fig:MSE1500} shows that the solutions of MPME are much better than those of SparSenSe when $k\gtrapprox n$. The reason is that the sensing indices corresponding to the $k$ largest elements of $\mathbf{w}^*$ cannot guarantee a  well-conditioned $\mathbf{\Psi}_k$ as shown in Fig. \ref{Fig:ConditionNO}.

\subsection{Comparison with FrameSense}

It is apparent from Fig. \ref{Fig:Performance} and Fig. \ref{Fig:ConditionNO} that FrameSense provides the worst results for the first two examples in which $\tilde{\mathbf{\Phi}}$ is not an equal-norm frame, i.e. the norms of the rows of $\tilde{\mathbf{\Phi}}$ are not equal. For these cases, minimization of the frame potential in \eqref{eq:FramePotential} will select the rows of $\tilde{\mathbf{\Phi}}$ with small norms to construct the observation matrix $\mathbf{\Phi}$. FrameSense prefers to drop the rows with large norms \cite{ranieri2014near}. It is easily found that
\begin{equation*}
\sum\nolimits_{i=1}^M \|\boldsymbol\varphi_{s_i}\|^2 = \mathbf{tr }(\mathbf{\Psi})=\sum\nolimits_{i=1}^n\lambda_i
\end{equation*}
Compared this equation with the MSE in \eqref{eq:MSE}, we conclude that small norms of the rows of the observation matrix $\mathbf{\Phi}$ lead to a large MSE of the estimated vector $\hat{\boldsymbol\alpha}$. From another perspective, we find that the smaller the norm of $\boldsymbol\varphi_{s_i}$ for $i=$ $1,..., M$, the smaller is the signal-to-noise ratio of the measurement model. Therefore, FrameSense is only suitable for the case where $\tilde{\mathbf{\Phi}}$ corresponds to an equal-norm frame.

\begin{figure*}[hbt]
    \centering
    \begin{subfigure}[b]{0.5\textwidth}
        \centering
        \includegraphics[width=\textwidth]{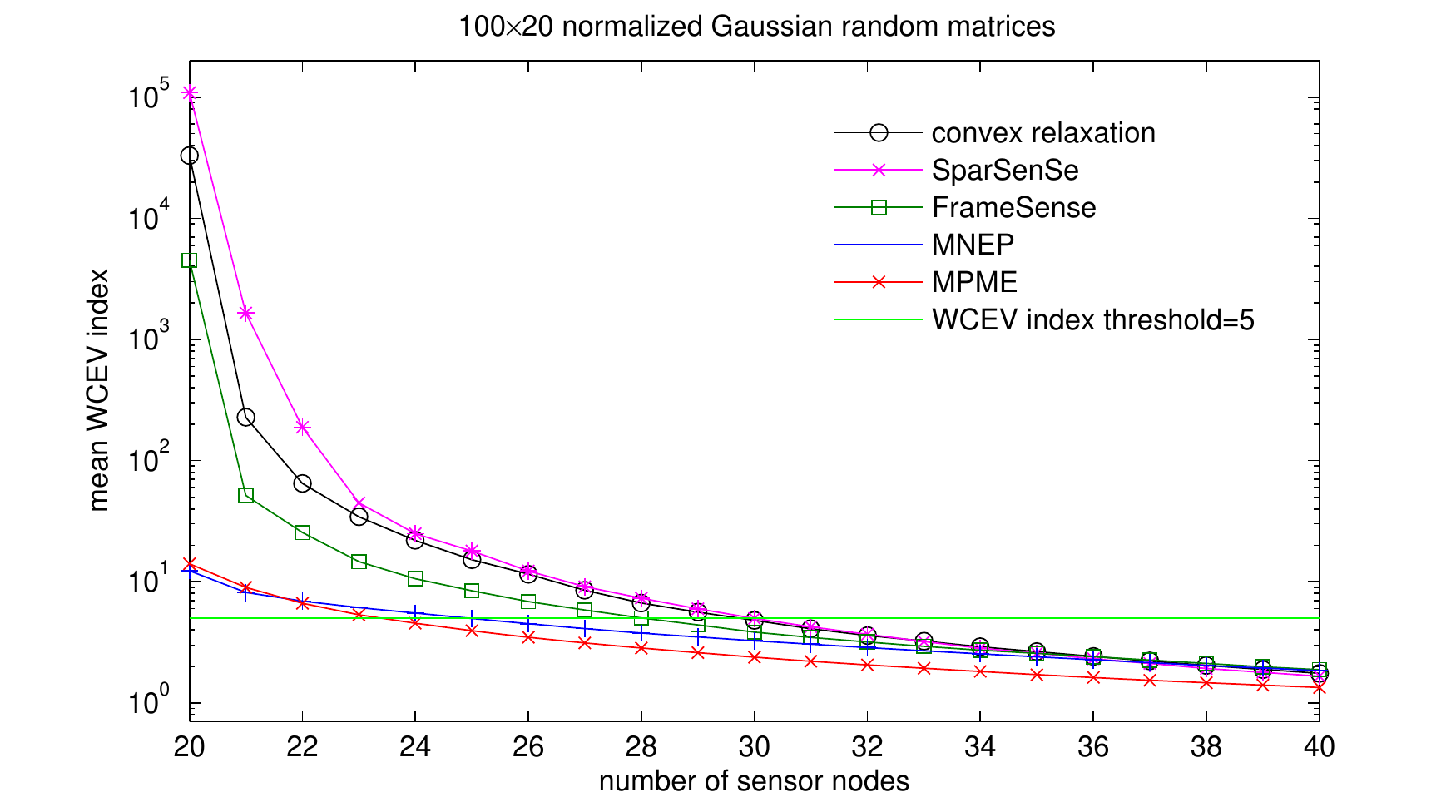}
    \end{subfigure}%
    \begin{subfigure}[b]{0.5\textwidth}
        \centering
        \includegraphics[width=\textwidth]{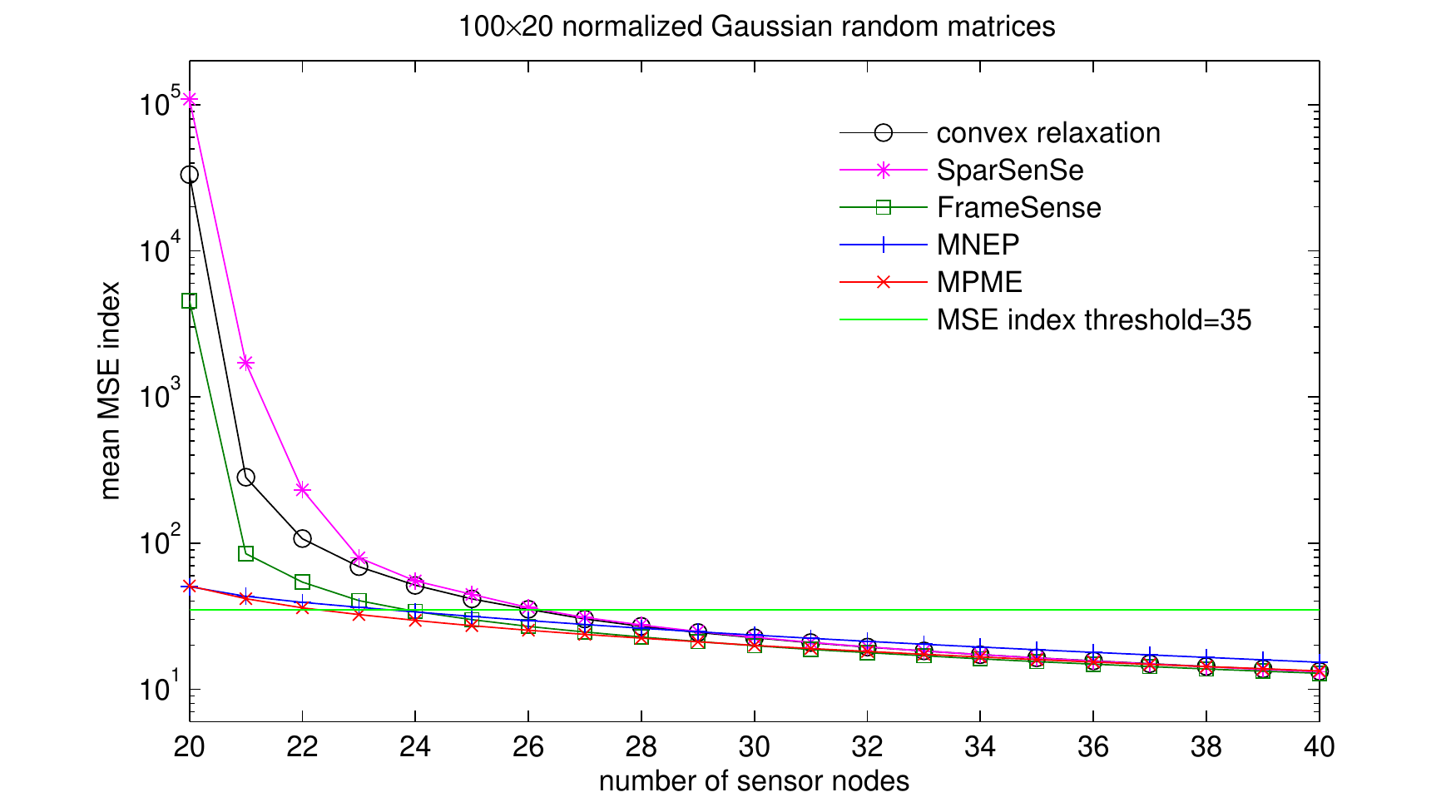}
    \end{subfigure}
    \caption{The mean WCEV index and mean MSE index of $\hat{\boldsymbol{\alpha}}$ estimated from 20 to 40 sensor observations for \emph{Example} \ref{example:normalizedGaussion}.}
    \label{Fig:NormalizedPerformance}
\end{figure*}

Actually, even if $\tilde{\mathbf{\Phi}}$ corresponds to an equal-norm frame, the proposed MPME algorithm still outperforms FrameSense, which will be illustrated by the following example.

\begin{example} \label{example:normalizedGaussion}
$\tilde{\mathbf{\Phi}}\in \mathbb{R}^{100\times 20}$
is a random matrix with normalized rows, whose $i$-th row $\boldsymbol\varphi_i^\mathrm{T}=\frac{\boldsymbol\phi_i^\mathrm{T}}{\|\boldsymbol\phi_i\|}$, and $\boldsymbol\phi_i\in \mathbb{R}^{20}$ is a random vector with independent components $\boldsymbol\phi_{ij}\sim \mathcal{N}(0,1)$. The variance of sensor noise $\sigma^2=1$.
\end{example}

The mean WCEV index and the mean MSE index of 200 Monte-Carlo run results are shown in Fig. \ref{Fig:NormalizedPerformance}. The two figures show that FrameSense outperforms the convex relaxation method and SparSenSe when the number of sensor nodes is small. The right figure shows that the five methods except MNEP provide almost the same mean MSE indices when the number of sensor nodes is large enough, which indicates the effectiveness of FrameSense in pursuing the minimum MSE.

However, like the convex relaxation method and SparSenSe,  in Fig. \ref{Fig:NormalizedPerformance}, both WCEV and MSE of FrameSense are much larger than those of MPME when the number of sensor nodes is small (e.g., $k=20$ or 21), which indicates that FrameSense cannot guarantee a well-conditioned $\mathbf{\Psi}_k$ when $k\gtrapprox n$.

Additionally, Fig. \ref{Fig:NormalizedPerformance} shows that MPME requires the least number of sensor nodes to meet the accuracy requirement, and that the required sensor nodes of FrameSense is less than those of the convex relaxation method and SparSenSe.

If $\tilde{\mathbf{\Phi}}$ corresponds to an equal-norm frame, the minimum frame potential of $\mathbf{\Phi}$ implies the minimum MSE, but the ``worst-out" strategy in FrameSense cannot find the optimal solution, and even cannot guarantee that $\mathbf{\Psi}$ is well-conditioned if the available sensor nodes is limited. The solution of FrameSense is near-optimal because of the sub-modularity of the cost function; however our MPME algorithm still outperforms FrameSense in the four examples. Hence, our future work may focus on exploring the reasons why MPME outperforms the near-optimal solution from a theoretical perspective.

\begin{figure*}[htb]
    \centering
    \begin{subfigure}[b]{0.5\textwidth}
        \centering
        \includegraphics[width=\textwidth]{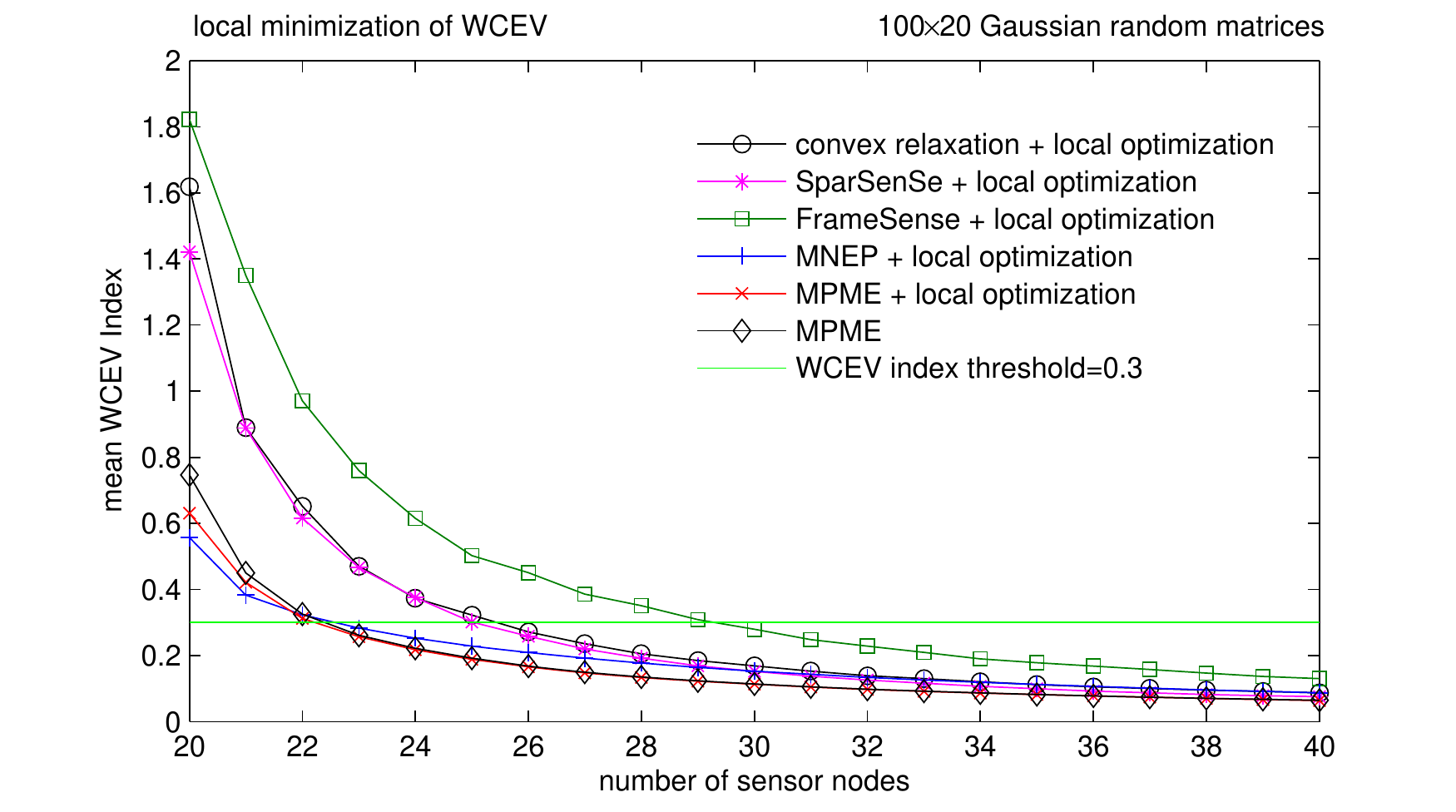}
    \end{subfigure}%
    \begin{subfigure}[b]{0.5\textwidth}
        \centering
        \includegraphics[width=\textwidth]{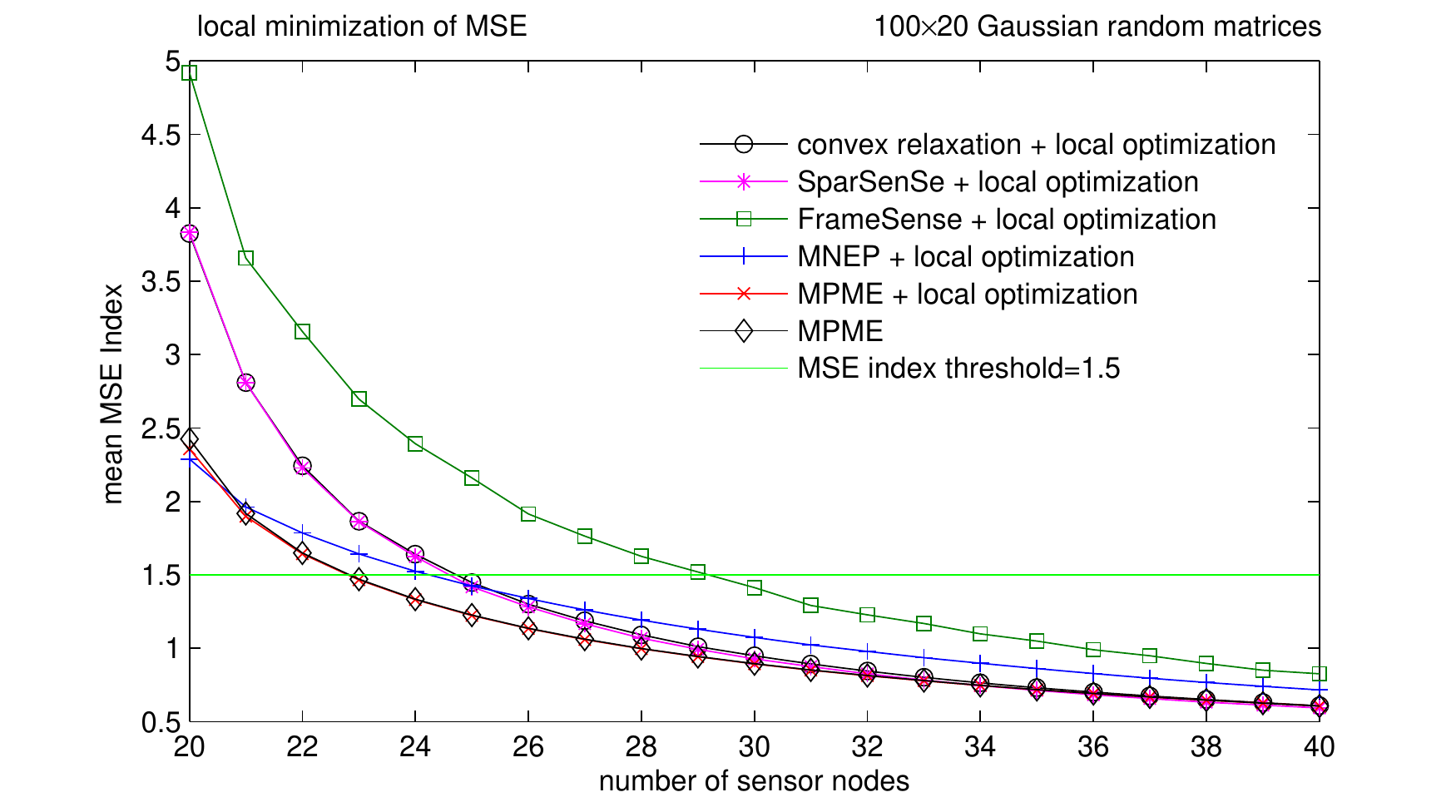}
    \end{subfigure}
    \begin{subfigure}[b]{0.5\textwidth}
    \centering
    \includegraphics[width=1\textwidth]{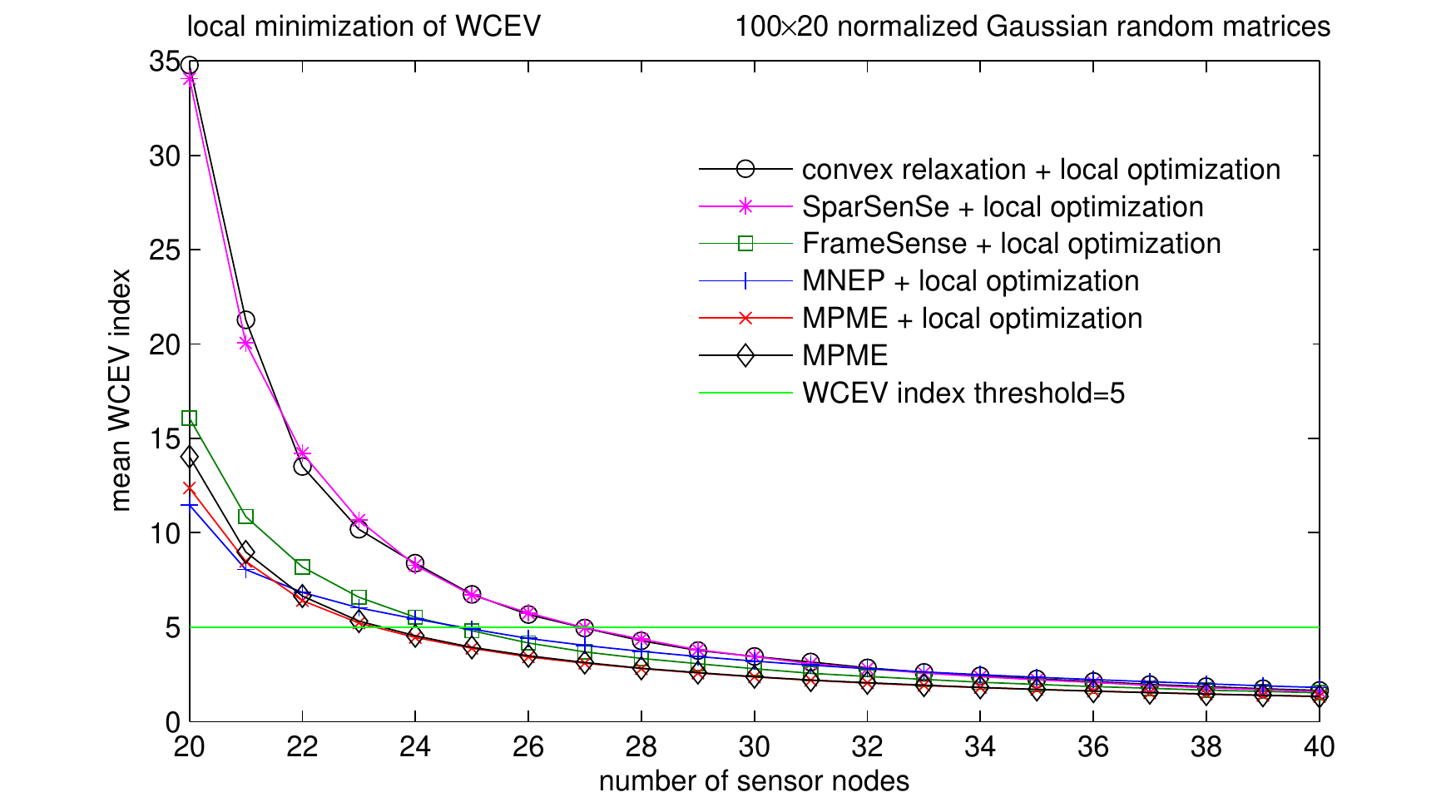}
    \end{subfigure}%
    \begin{subfigure}[b]{0.5\textwidth}
        \centering
        \includegraphics[width=\textwidth]{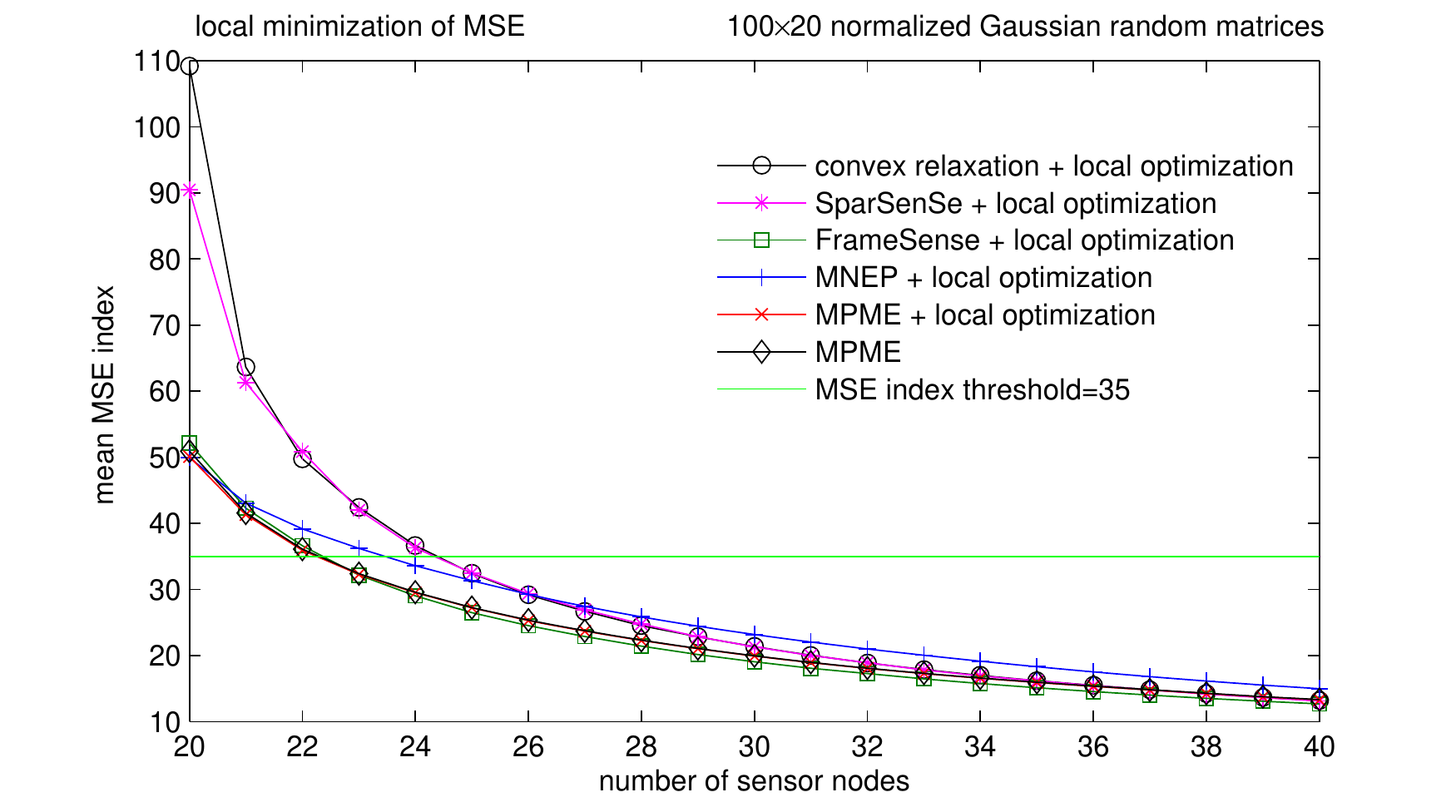}
    \end{subfigure}
    \caption{The performance comparison between MPME and five methods with local optimization for \emph{Example} \ref{example:generalGaussion} and \emph{Example} \ref{example:normalizedGaussion}.}
    \label{Fig:LocalPerformance}
\end{figure*}

\subsection{Local optimization}

The four examples show that the current methods are not suitable for the cases that the number of sensors is small. This drawback can be overcome by a computationally expensive technique, i.e. the so called \emph{local optimization} technique \cite{joshi2009sensor}.

\begin{definition} [Local optimization]
For a given set of sensing locations $\mathcal{S}$, exchange one-at-a-time all the sensing location in $\mathcal{S}$ with each available candidate location in $\mathcal{N}/\mathcal{S}$ to re-locate the sensor nodes at new position that can further reduce one criterion of interest (e.g., MSE or WCEV) until there is no further decrease.
\end{definition}

This technique is similar with Fedorov's exchange algorithm \cite{miller1994algorithm}, and Wynn's algorithm \cite{wynn1972results}. It has also been discussed in \cite{Yildirim2009efficient, joshi2009sensor}.
For any results of local optimization, replacing any selected sensing location by any unselected one cannot improve the solution, which is called \emph{2-opt}.

We apply the local optimization technique to the solutions obtained from the five methods for the four examples. Fig. \ref{Fig:LocalPerformance} shows the mean WCEV indices and the mean MSE indices of the improved sensor configurations, together with those directly obtained from MPME, i.e. without local optimization.
In Fig. \ref{Fig:LocalPerformance} the solutions of the convex relaxation method, SparSenSe, and FrameSense are remarkably improved by the local optimization, especially when the number of sensor nodes is small. However, the solutions of MPME almost have no improvement. Nevertheless, the two top figures show that the solution of MPME without local optimization still outperforms all the other solutions with local optimization in terms of both indices.

The bottom right figure shows that with local optimization, the solution of FrameSense for normalized Gaussian random matrices are sightly better than the solution of MPME without local optimization. However, the local optimization is computationally very expensive, and the MPME provides the best result amongst the five methods for all the other cases. How the local optimization affects a given sensor configuration and which types of sensor configuration can be greatly improved by local optimization are still open problems.

Additionally, Fig. \ref{Fig:LocalPerformance} shows that for all cases, to meet the accuracy requirement, the solutions of MPME without local optimization require the least number of sensor nodes. From this perspective, MPME without local optimization outperforms the state-of-the-art with local optimization.

\begin{table*}[tbh]
\caption{The computational effort of the five sensor placement methods}
\begin{center}
\begin{tabular}{ccccc}
\hline\hline
Convex Relaxation & SparSenSe & FrameSense & MNEP & MPME\\
\hline
$O(i_\mathrm{c}N^3)$ & $O(i_sN^3)$ & $O(N^3)$&$O(NMn^3)$ &$O(NMn^2)$\\
\hline\hline
\end{tabular}

\label{Table:ComputationalCost}
\end{center}
\end{table*}

\section{computational cost of the MPME algorithm}

In this section, we compare the computational cost of MPME with that of the state-of-the-art.

The computational effort of the convex relaxation method is $O(i_\mathrm{c}N^3)$ \cite{joshi2009sensor}. The convex optimization problem is solved using the interior-point method and $i_\mathrm{c}$ is the iteration number Typically, the iteration number is of a few tens \cite{joshi2009sensor}.

%The main computational effort of the SparSenSe method is to solve the LMI problem \eqref{eq:SparSenSe}. The computational cost of solving one LMI problem is $O(\tau N^3)$ \cite{gahinet1995lmi, Tarbouriech2011Stability} where $\tau$ is the row size of the LMI system, and $N$ is the total number of scalar decision variables. The row size is the number of the lines of the LMI constraints. For more detail about the row size of LMI systems, see page 132 and the examples on page 133 of \cite{Tarbouriech2011Stability}. For the LMI problem \eqref{eq:SparSenSe}, $\tau=(n+1)^2+N$. Hence, the computational effort of SparSenSe is $O\left(N^4+N^3(n+1)^2\right)$. If $n\ll N$, the computational effort is $O(N^4)$.

Similar to the convex relaxation method, the computational effort of SparSenSe is $O(i_sN^3)$ where $i_s$ is the iteration number of solving the convex optimization problem \eqref{eq:SparSenSe}.

When using FrameSense, $N-M$ rows are removed from $\tilde{\mathbf{\Phi}}$. It costs $O\left((N-k+1)^2\right)$ to determine the $k$-th removed row. Since $M\ll N$, the total cost of FrameSense is $O(\sum_{i=1}^Ni^2-\sum_{i=1}^Mi^2)=O(N^3)$.

Finding the eigenvalues of $\mathbf{\Psi}_k$ costs $O(n^3)$ operations. When we find the $k$-th sensing location via MNEP, the main computational cost is to solve the minimum nonzero eigenvalue maximization problem in which $N-k+1$ eigenvalue problems are solved. The computation cost is $O\left(Nn^3\right)$. Therefore, to determine all the $M$ sensing locations via MNEP, the total computational effort is $O(NMn^3)$.

To determine the $k$-th sensing location via MPME, the main computational cost is attributed to the optimization problem $\hat s_k = \underset{i\in\mathcal N \setminus \mathcal S} {\arg\,\max}~\|\mathbf{P}_{k-1}\boldsymbol{\varphi}_i\|_2$ which costs $O\left((N-k+1)n^2\right)$. Hence, finding all the $M$ sensing locations costs $O(NMn^2)$.

\begin{figure}[h]
\centering
\includegraphics[width=0.48\textwidth]{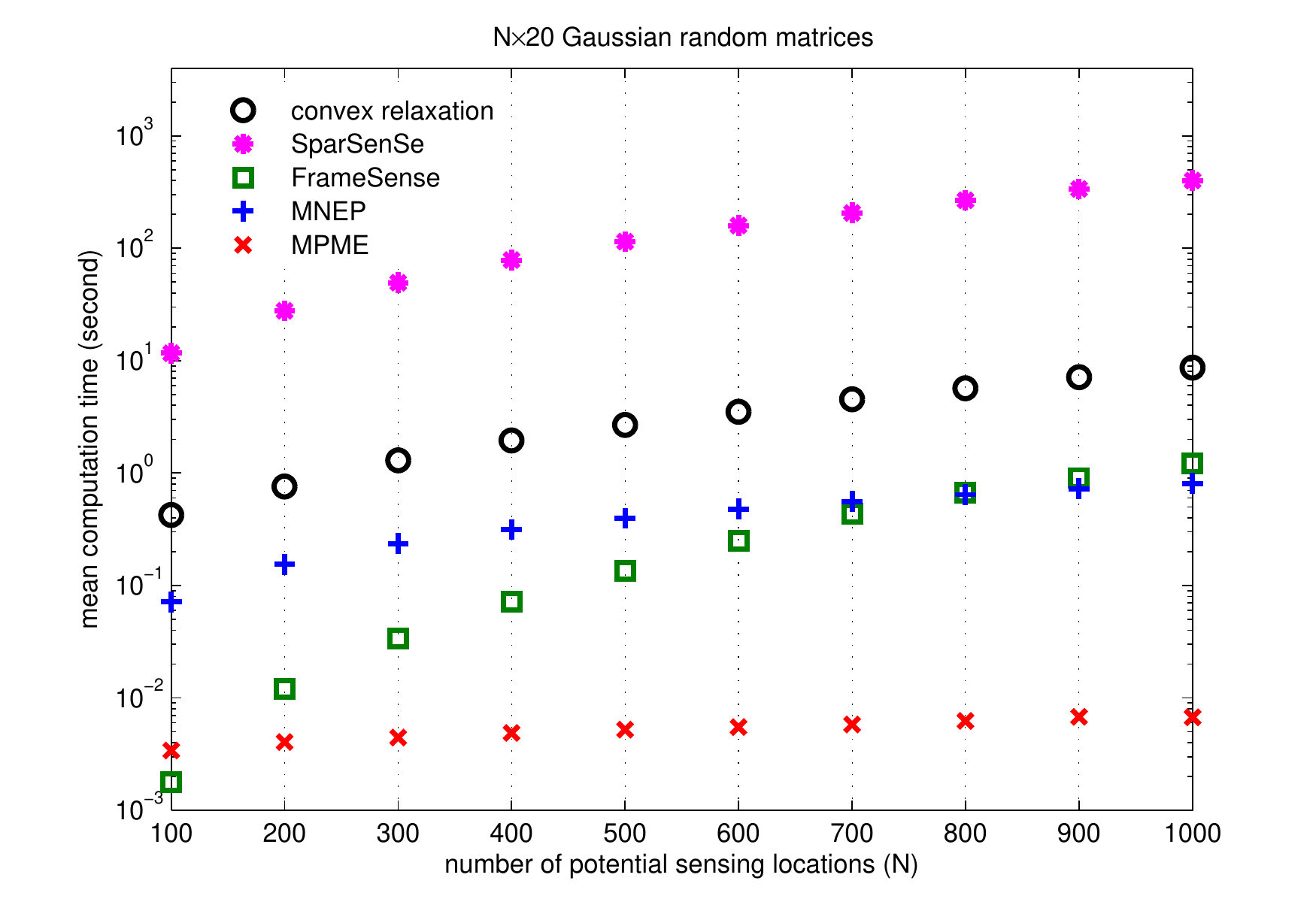}
\caption{The mean computation time of the five sensor placement algorithms for $N\times20$ Gaussian random matrices where $N$ is from 100 to 1000. In the simulations, the number of available sensor nodes, i.e. $M$,  is set as 20 for the convex relaxation method, FrameSenSe, MNEP and MPME. The computation time is estimated by the MATLAB tic-toc commands. The simulation is run in a laptop with a 2.4GHz Intel i3-3110M processor. The mean computation time is the mean value of 50 different simulations, and for each simulation, $\tilde{\mathbf{\Phi}}\in\mathbb{R}^{N\times20}$ with independent entries $\varphi_{ij}\sim \mathcal{N}(0,1)$.}
\label{Fig:CompuationalTime}
\end{figure}

We summarize the computational efforts of the five methods in Table \ref{Table:ComputationalCost}. The mean computation time of 50 simulation run results are shown in Fig. \ref{Fig:CompuationalTime}. Both Table \ref{Table:ComputationalCost} and Fig. \ref{Fig:CompuationalTime} show that MPME is computationally the most efficient one amongst the five algorithms if $N\gg M\geq n$.

\section{Conclusions}

Sensor placement for linear inverse problems is an interesting but challenging combinatorial problem. The optimal solution can be solved by the exhaustive search and branch-and-bound methods \cite{lawler1966branch, welch1982branch}, but the methods are both impractical due to the extremely expensive computational cost, especially for some large scale problems. Therefore, in the last decade, many works have focused on finding an effective suboptimal solution via computationally efficient algorithms. To the best of our knowledge, the proposed MPME algorithm is computationally one of the most efficient sensor placement algorithms.

Our proposed MNEP and MPME algorithms select the sensing locations one-by-one. In this way, the minimum number of the required sensor nodes can be readily determined. Different with many popular methods, MNEP and MPME can guarantee that the dual observation matrix is well-conditioned when the number of sensor nodes is small and even near the dimension of the estimated vector.

The sufficient and necessary condition of meeting the requirement on WCEV is shown to be that the square summation of the projections of all selected observation vectors onto any non-trivial subspace of $\mathbb{R}^n$ is large enough. The proposed MPME algorithm determines each sensing location by maximizing the projection of its observation vector onto the subspace onto which the square summation of the projections of all selected observation vectors is minimum.

We perform Monte-Carlo simulations to compare the MNEP and MPME algorithms with the convex relaxation method \cite{joshi2009sensor}, SparSenSe \cite{jamali2014sparsity}, and FrameSense \cite{ranieri2014near}. Based on the simulation results, we conclude that amongst the five methods:
\begin{itemize}
  \item To meet the accuracy requirement, the solution of MPME requires the least number of sensor nodes;

  \item The MPME algorithm provides the best solution in the sense of minimum WCEV or minimum MSE, especially when the number of sensor nodes used is small;

  \item MNEP and MPME work well when the number of available sensor nodes is very limited, while the state-of-the-art cannot;

  \item For the general cases, the solution of the MPME without local optimization is even better than those of the state-of-the-art with local optimization.
  %\item The MNEP algorithm outperforms the state-of-the art when the number of sensor nodes used is small.
\end{itemize}

To encourage future works, we provide all the Matlab code used in this paper, which can be found from IEEE Xplore or  \url{https://github.com/CJiang01/SensorPlacement.git}.

\section*{Appendix A\\Proof of Theorem \ref{Theorem:Eigenvalue&Projection}}

\begin{IEEEproof}
The spectrum decomposition of $\mathbf{\Psi}_k$ is
\begin{equation}
\mathbf{\Psi}_k={\mathbf{\Phi}}_k^\mathrm{T}{\mathbf{\Phi}}_k=\mathbf{U}^{(k)}\mathbf{\Lambda}_k(\mathbf{U}^{(k)})^\mathrm{T}
\label{eq:SchurDecomposition}
\end{equation}
where $\mathbf{U}^{(k)}=[\mathbf{u}_1^{(k)}, \mathbf{u}_2^{(k)},..., \mathbf{u}_n^{(k)}]$ is an orthonormal matrix, and $\mathbf{\Lambda}_k$ is a diagonal matrix whose diagonal entry $\mathbf{\Lambda}_{ii}^{(k)}=\lambda_i^{(k)}$.
Then, we can obtain
\begin{equation*}
\mathbf{\Lambda}_k=(\mathbf{U}^{(k)})^\mathrm{T}{\mathbf{\Phi}}_k^\mathrm{T}{\mathbf{\Phi}}_k\mathbf{U}^{(k)}
\end{equation*}
from which \eqref{eq:eigenvalueEQ} can be directly found.
\end{IEEEproof}

\section*{Appendix B\\Proof of Theorem \ref{Theorem:S_and_N}}

To proof Theorem \ref{Theorem:S_and_N}, we need the following lemma.
\begin{lemma}[Courant-Fischer Minimax Theorem]
\label{Theorem:CFMT}
If $\mathbf{A}\in\mathbb{R}^{n\times n}$ is symmetric, then for $i = 1:n$,
\begin{equation}
\lambda_i(\mathbf{A}) =  \underset{\mathrm{\mathbf{dim}}(\mathbb{U})=i}{\mathrm{max}} \;\;  \underset{\mathbf{0}\neq \mathbf{x}\in\mathbb{{U}}}{\mathrm{min}} \frac{\mathbf{x}^\mathrm{T}\mathbf{A}\mathbf{x}}{\mathbf{x}^\mathrm{T}\mathbf{x}} \label{eq:eigenvalueCFM}
\end{equation}
\end{lemma}

\begin{IEEEproof}
See the proof of Theorem 8.1.2 in \cite{golub2012matrix}.
\end{IEEEproof}

Next, we prove Theorem \ref{Theorem:S_and_N}.
\begin{IEEEproof}
Since $ \mathbf{x}^\mathrm{T}\mathbf{\Psi}_k\mathbf{x}= \mathbf{x}^\mathrm{T}\mathbf{\Phi}_k^\mathrm{T}\mathbf{\Phi}_k\mathbf{x} =\|\mathbf{\Phi}_k\mathbf{x}\|_2^2$, from Lemma \ref{Theorem:CFMT} we have
\begin{equation}
\lambda_n^{(k)} = \underset{\|\mathbf{x}\|_2=1}{\mathrm{min}} \mathbf{x}^\mathrm{T}\mathbf{\Psi}_k\mathbf{x} =\underset{\|\mathbf{x}\|_2=1}{\mathrm{min}} \|\mathbf{\Phi}_k\mathbf{x}\|_2^2 \label{eq:MinRayleighQuotient}
\end{equation}
Then, from \eqref{eq:eigenvalueEQ} and \eqref{eq:MinRayleighQuotient}, we can obtain \eqref{eq:EigenvectorDirection}.

Next, we show the sufficient and necessary condition of $\lambda_n^{(k)}>\gamma$ is that for any nonzero normalized vector $\mathbf{x}\in\mathbb{R}^n$, $\|\mathbf{\Phi}_k\mathbf{x}\|_2^2>\gamma$.

\emph{Sufficiency:} Considering \eqref{eq:eigenvalueEQ}, if $\|\mathbf{\Phi}_k\mathbf{x}\|_2^2>\gamma$ for any nonzero normalized vector $\mathbf{x}\in\mathbb{R}^n$, we can obtain $\lambda_n^{(k)}>\gamma$.

\emph{Necessity:} Since $\lambda_n^{(k)}>\gamma$, considering \eqref{eq:MinRayleighQuotient} we conclude that $\|\mathbf{\Phi}_k\mathbf{x}\|_2^2>\gamma$ for any nonzero normalized vector $\mathbf{x}\in\mathbb{R}^n$.
\end{IEEEproof}

\section*{Appendix C\\Proof of Theorem \ref{Theorem:PCorrelated}}

Considering \eqref{eq:zdifinition} and \eqref{eq:SchurDecomposition}, we can obtain another description of $\mathbf{\Psi}_k$, i.e.
\begin{equation}
\mathbf{\Psi}_k\!\!=\!{\mathbf{\Phi}}_{k\!-\!1}^\mathrm{T}\!{\mathbf{\Phi}}_{k\!-\!1}\!+\boldsymbol{\varphi}_{s_{k}}\!\boldsymbol{\varphi}_{s_{k}}^\mathrm{T} \!=\!\mathbf{U}^{(k-\!1)}\!(\!\mathbf{\Lambda}_{k-\!1}\!+\mathbf{zz}^\mathrm{T}\!)(\!\mathbf{U}^{(k-\!1)}\!)^\mathrm{T}
\label{eq:SchurDecApproxExpan}
\end{equation}
Let $\hat {\mathbf{\Lambda}}_k = \mathbf{\Lambda}_{k-1}+\mathbf{zz}^\mathrm{T}$. From \eqref{eq:SchurDecomposition} and \eqref{eq:SchurDecApproxExpan}, we find that $\mathbf{\Psi}_k$, $\mathbf{\Lambda}_k$, and $\hat{\mathbf{\Lambda}}_k$ are mutually \emph{similar}, which implies that they share the same eigenvalues.

To proof {Theorem} \ref{Theorem:PCorrelated}, we need the following four lemmas.

\begin{lemma}\label{lemma:zequals0}
If $z_i=0$, $\lambda_i^{(k-1)}$ is an eigenvalue of $\hat{\mathbf{\Lambda}}_{k}$, and the corresponding  eigenvector is $\mathbf{e}_i$
\end{lemma}
\begin{IEEEproof}
As $z_i=0$, we can obtain
  \begin{equation*}
  \hat{\mathbf{\Lambda}}_{k}\mathbf{e}_i=\mathbf{\Lambda}_{k-1}\mathbf{e}_i+\mathbf{z}\mathbf{z}^\mathrm{T}\mathbf{e}_i= \lambda_i^{(k-1)}\mathbf{e}_i
  \end{equation*}
It is clear that $\lambda_i^{(k-1)}$ is an eigenvalue of $\hat{\mathbf{\Lambda}}_k$ and $\mathbf{e}_i$ is the corresponding eigenvector.
\end{IEEEproof}

\begin{lemma}\label{lemma:pruned}
For all $i\in\mathcal{I}=\{i|1\leq i \leq n, z_i\neq 0\}$, denote by $\mathbf{v}_i$ the eigenvector of $\hat {\mathbf{\Lambda}}_{k}$ associated with $\lambda_i^{(k)}$. For all $j\in \mathcal{J}= \{j|1\leq j \leq n, z_j=0 \}$, removing the $j$-th row and $j$-th column of $\mathbf{\Lambda}_{k-1}$ and $\hat{\mathbf{\Lambda}}_k$ yields $\tilde{\mathbf{\Lambda}}_{k-1}$ and $\tilde{\mathbf{\Lambda}}_k$, respectively, and removing the $j$-th entry of $\mathbf{v}_i$ and $\mathbf{z}$ yields $\tilde{\mathbf{v}}_i$ and $\tilde{\mathbf{z}}$, respectively. Then,
\begin{equation}
\tilde{\mathbf{\Lambda}}_{k}\tilde{\mathbf{v}}_i = \tilde{\mathbf{\Lambda}}_{k-1}\tilde{\mathbf{v}}_i+ \tilde{\mathbf{z}}\tilde{\mathbf{z}}^\mathrm{T}\tilde{\mathbf{v}}_i = \lambda_i^{(k)}\tilde{\mathbf{v}}_i\label{eq:pruned}
\end{equation}
\end{lemma}
\begin{IEEEproof}
For all $j\in \mathcal{J}$, removing the $j$-th row of the following equation directly yields \eqref{eq:pruned}.
\begin{equation*}
{\mathbf{\Lambda}}_{k}{\mathbf{v}}_i =\mathbf{\Lambda}_{k-1}\mathbf{v}_i+\mathbf{z}\left(\mathbf{z}^\mathrm{T}\mathbf{v}_i\right)= \lambda_i^{(k)}\mathbf{v}_i
\end{equation*}
\end{IEEEproof}

\begin{lemma} \label{lemma: multiplicity}
If $\lambda_i^{(k-1)}=\lambda_{i+\mu_i-1}^{(k-1)}$ is an eigenvalue of $\mathbf{\Lambda}_{k-1}$ with multiplicity $\mu_i$ and $\sum_{j=i}^{i+\mu_i-1} z_j^2 \neq 0$, then $\lambda_i^{(k-1)}$ is an eigenvalue of $\hat{\mathbf{\Lambda}}_{k}$ with multiplicity $\mu_i-1$.
\end{lemma}
\begin{IEEEproof}
Since $\sum_{j=i}^{i+\mu_i-1} z_j^2 \neq 0$, we can obtain
\begin{eqnarray*}
\mathrm{\mathbf{rank}}(\hat{\mathbf{\Lambda}}_{k}-\lambda_{i}^{(k-1)}\mathbf{I})
&=& \mathrm{\mathbf{dim}}~\mathrm{\mathbf{span}}(\hat{\mathbf{\Lambda}}_{k}-\lambda_{i}^{(k-1)}\mathbf{I})  \\
&=& \mathrm{\mathbf{dim}}~\mathrm{\mathbf{span}}([\mathbf{\Lambda}_{k-1}-\lambda_{i}^{(k-1)}\mathbf{I}~~~\mathbf{z}]) \notag \\
&=& n-\mu_{i}+1
\end{eqnarray*}

where for any matrix $\mathbf{A}$, $\mathrm{\mathbf{dim}}~\mathrm{\mathbf{span}}(\mathbf{A})$ represents the dimension of a linear space spanned by all the column vectors of $\mathbf{A}$.   Therefore, $\lambda_{i}^{(k-1)}$ is an eigenvalue of $\hat{\mathbf{\Lambda}}_{k}$ with multiplicity $\mu_{i}-1$.
\end{IEEEproof}

\begin{lemma} \label{lemma: multipleEigen}
 If $\lambda_n^{(k-1)}$ is a multiple eigenvalue of $\mathbf{\Lambda}_{k-1}$ with multiplicity $\mu_n$ and $\sum_{j=n-\mu_{n}+1}^{n} z_j^2 \neq 0$, then $\lambda_{n-\mu_n+1}^{(k)}\neq \lambda_i^{(k-1)}$ for all $i$ satisfying $z_i\neq 0$.
\end{lemma}
\begin{IEEEproof}
It follows from {Theorem \ref{Theorem:sequence}} that
\begin{equation}
\lambda_i^{(k-1)}\leq \lambda_i^{(k)} \leq \lambda_{i-1}^{(k-1)} \quad \mathrm{for}\; \mathrm{all}\; 1<i\leq n \label{eq:lambdakk}
\end{equation}
Hence,
\begin{equation}
\lambda_{n-\mu_n+1}^{(k-1)}\leq \lambda_{n-\mu_n+1}^{(k)} \leq \lambda_{n-\mu_n}^{(k-1)}  \label{eq:lambdak&k-1}
\end{equation}
Then, considering Lemma \ref{lemma: multiplicity}, we can obtain
\begin{equation}
\lambda_{n-\mu_n+1}^{(k)}>\lambda_{n-\mu_n+1}^{(k-1)} = \lambda_{n}^{(k-1)} =\lambda_{n-\mu_n+2}^{(k)}=\lambda_n^{(k)} \label{eq:lambda_mul}
\end{equation}

Denote the multiplicity of $\lambda_{n-\mu_n}^{(k-1)}$ w.r.t. $\mathbf{\Lambda}_{k-1}$ by $\mu$, where $\mu\geq 1$. Hence,
\begin{equation}
\lambda_{n-\mu_n-\mu}^{(k-1)}> \lambda_{n-\mu_n-\mu+1}^{(k-1)} = \lambda_{n-\mu_n-\mu+2}^{(k-1)}=...=\lambda_{n-\mu_n}^{(k-1)} \label{eq:multiplicityk-1}
\end{equation}

If $\sum_{i=n-\mu_{n}-\mu+1}^{n-\mu_n} z_i^2 \neq 0$, considering \emph{Lemma \ref{lemma: multiplicity}}, $\lambda_{n-\mu_n}^{(k-1)}$ is an eigenvalue of $\hat{\mathbf{\Lambda}}_k$ with multiplicity $\mu-1$. Therefore, if
\begin{equation}
\lambda_{n-\mu_n}^{(k-1)}= \lambda_{n-\mu_n+1}^{(k)}
\end{equation}
the multiplicity of $\lambda_{n-\mu_n+1}^{(k)}$ w.r.t. $\hat{\mathbf{\Lambda}}_k$ is $\mu-1$ and
\begin{equation}
\lambda_{n-\mu_n+1}^{(k)}= \lambda_{n-\mu_n}^{(k)} =...= \lambda_{n-\mu_n-\mu+3}^{(k)}<\lambda_{n-\mu_n-\mu+2}^{(k)} \label{eq:multiplicityk}
\end{equation}
From \eqref{eq:multiplicityk-1}-\eqref{eq:multiplicityk}, we can obtain $\lambda_{n-\mu_n-\mu+1}^{(k-1)}<\lambda_{n-\mu_n-\mu+2}^{(k)}$, which obviously contradict with \eqref{eq:lambdakk}. Hence, $\lambda_{n-\mu_n+1}^{(k)}\neq\lambda_{n-\mu_n}^{(k-1)}$.

Consequently, if $\sum_{i=n-\mu_{n}-\mu+1}^{n-\mu_n} z_i^2 \neq 0$, considering \eqref{eq:lambdak&k-1} and \eqref{eq:lambda_mul} we can obtain
\begin{equation*}
\lambda_{n-\mu_n+1}^{(k-1)}<\lambda_{n-\mu_n+1}^{(k)}<\lambda_{n-\mu_n}^{(k-1)}\leq \lambda_j^{(k-1)},~~j<n-\mu_n
\end{equation*}
else ($\sum_{i=n-\mu_{n}-\mu+1}^{n-\mu_n} z_i^2 = 0$)
\begin{equation*}
\lambda_{n-\mu_n+1}^{(k-1)}<\lambda_{n-\mu_n+1}^{(k)}\leq\lambda_{n-\mu_n}^{(k-1)}\leq \lambda_j^{(k-1)},~~j<n-\mu_n
\end{equation*}
which implies this lemma.
\end{IEEEproof}

Next, leveraging the four Lemmas, we are ready to prove Theorem \ref{Theorem:PCorrelated}.

\begin{IEEEproof} Let $\lambda_n^{(k-1)}$ be the minimum eigenvalue of $\mathbf{\Psi}_{k-1}$ ($\mathbf{\Lambda}_{k-1}$) with multiplicity $\mu_n(\geq 1)$. If $k<n$, $\lambda_n^{(k-1)}=0$ and $\mu_n=n-k+1$. From {Lemma \ref{lemma: multiplicity}}, we can directly obtain \eqref{eq:multi-Eigenvalue} and \eqref{eq:multi-Eigenvalue_n}.

If $\sum_{i=n-\mu_n+1}^{n}z_{i}^2=0$, it is obvious that $\zeta_k=0$. According to \emph{Lemma \ref{lemma:zequals0}}, we can easily obtain \eqref{eq:MinmultiEigenvalue} and \eqref{eq:MinEigenvalue}.

If $\sum_{i=n-\mu_n+1}^{n}z_{i}^2\neq0$, according to \emph{Lemma \ref{lemma:pruned}}, we can obtain
\begin{equation}
 (\tilde{\mathbf{\Lambda}}_{k-1}-\lambda_{n-\mu_n+1}^{(k)}\mathbf{I})\tilde{\mathbf{v}}_{n-\mu_n+1}+ \tilde{\mathbf{z}}(\tilde{\mathbf{z}}^\mathrm{T}\tilde{\mathbf{v}}_{n-\mu_n+1})=0 \label{eq:prunedEigenk<n}
\end{equation}
\emph{Lemma \ref{lemma: multipleEigen}} can guarantee that $\tilde{\mathbf{\Lambda}}_{k-1}-\lambda_{n-\mu_n+1}^{(k)}\mathbf{I}$ is nonsingular, and therefore from \eqref{eq:prunedEigenk<n}, we find that $\tilde{\mathbf{z}}^\mathrm{T}\tilde{\mathbf{v}}_{n-\mu_n+1}\neq 0$. Then, left multiplying $\tilde{\mathbf{z}}^\mathrm{T}(\tilde{\mathbf{\Lambda}}_{k-1}-\lambda_{n-\mu_n+1}^{(k)}\mathbf{I})^{-1}$ to both sides of \eqref{eq:prunedEigenk<n} yields

\begin{equation*}
\tilde{\mathbf{z}}^\mathrm{T}\tilde{\mathbf{v}}_{n-\mu_n+1}\left(1+\tilde{\mathbf{z}}^\mathrm{T} (\tilde{\mathbf{\Lambda}}_{k-1}-\lambda_{n-\mu_n+1}^{(k)}\mathbf{I})^{-1}\tilde{\mathbf{z}}\right)=0
\end{equation*}
Hence,
\begin{equation*}
1+\tilde{\mathbf{z}}^\mathrm{T} (\tilde{\mathbf{\Lambda}}_{k-1}-\lambda_{n-\mu_n+1}^{(k)}\mathbf{I})^{-1}\tilde{\mathbf{z}}=0
\end{equation*}
from which we can directly obtain \eqref{eq:MinEigenvalue}. If $k\leq n$, $\lambda_{k}^{(k-1)}=0$ and $\mu_n=n-k+1$, substituting them into \eqref{eq:MinEigenvalue} yields \eqref{eq:MinmultiEigenvalue}.

It is clear that equation \eqref{eq:MinEigenvalue} is a general description of \eqref{eq:MinmultiEigenvalue}. From \eqref{eq:MinEigenvalue}, we can obtain

\begin{equation}
\zeta_k \!=\! \lambda_{n-\!\mu_n\!+1}^{(k)}\!-\!\lambda_{n-\!\mu_n\!+1}^{(k-\!1)}+\!\!\! \sum\limits_{i=1,z_i\neq 0}^{n-\mu_n}\!\!\!\frac{(\lambda_{n-\!\mu_n\!+1}^{(k)}\!-\!\lambda_{n-\!\mu_n\!+1}^{(k-\!1)})z_i^2}{\lambda_i^{(k-1)}-\lambda_{n-\mu_n+1}^{(k)}} \label{eq:projection}
\end{equation}

Taking the derivative of $\zeta_k$ w.r.t. $\lambda_{n-\!\mu_n\!+1}^{(k)}$, we can obtain that
\begin{equation*}
\frac{\mathrm{d}\zeta_k}{\mathrm{d}\lambda_{n-\!\mu_n\!+1}^{(k)}} = 1+\sum\limits_{i=1,z_i\neq 0}^{n-\mu_n}\frac{(\lambda_{i}^{(k-1)}-\lambda_{n-\!\mu_n\!+1}^{(k-1)})z_i^2}{(\lambda_i^{(k-1)}-\lambda_{n-\!\mu_n\!+1}^{(k)})^2}
\end{equation*}
Since $\lambda_i^{(k-1)}\geq \lambda_{n-\mu_n+1}^{(k-1)}$ for all $i\leq n-\mu_n$, it is obvious that  ${\mathrm{d}\zeta_k}/{\mathrm{d}\lambda_n^{(k)}}\geq 1$. Therefore, $0<{\mathrm{d}\lambda_{n-\mu_n+1}^{(k)}}/{\mathrm{d}\zeta_k}\leq 1$, which implies that $\lambda_{n-\mu_n+1}^{(k)}$ is monotonically strictly increasing w.r.t. $\zeta_k$, i.e.
\begin{equation}
\zeta_k \leftrightarrow\lambda_{n-\mu_n+1}^{(k)}\label{eq:zeta}
\end{equation}
where `$a\leftrightarrow b$' means that $b$ is monotonically increasing w.r.t. $a$.

Taking derivative of both sides of \eqref{eq:projection} w.r.t. $\lambda_{n-\mu_n}^{(k-1)}$ and with some operations, we can obtain
\begin{equation*}
\frac{\mathrm{d}\lambda_{n-\mu_n+1}^{(k)}}{\mathrm{d}\lambda_{n-\mu_n}^{(k-1)}}=\frac{\lambda_{n-\mu_n+1}^{(k)}z_{n-\mu_n}^2}
{(\lambda_{n-\mu_n}^{(k-1)}-\lambda_{n-\mu_n+1}^{(k)})^2(1+ x)}
\end{equation*}
where
\begin{equation*}
x=\sum\limits_{i=1,z_i\neq 0}^{n-\mu_n}\frac{\lambda_{i}^{(k-1)}z_i^2}{(\lambda_i^{(k-1)}-\lambda_{n-\mu_n+1}^{(k)})^2}
\end{equation*}
Since $\lambda_{n-\mu_n+1}^{(k)}>0$ and $\lambda_i^{(k-1)}>0$ for all $i\leq n-\mu_n$, we can obtain that ${\mathrm{d}\lambda_{n-\mu_n+1}^{(k)}}/{\mathrm{d}\lambda_{n-\mu_n}^{(k-1)}}\geq 0$, which implies that $\lambda_{n-\mu_n+1}^{(k)}$ is monotonically increasing w.r.t. $\lambda_{n-\mu_n}^{(k-1)}$, i.e.
\begin{equation}
\lambda_{n-\mu_n}^{(k-1)}\leftrightarrow \lambda_{n-\mu_n+1}^{(k)} \label{eq:lambdakk-1_leqn}
\end{equation}
If $k\leq n$, we have $\mu_n=n-k+1$ and therefore, $\lambda_{k}^{(k)}$ is monotonically increasing w.r.t. $\lambda_{k-1}^{(k-1)}$ for all $k\leq n$, i.e.
\begin{equation}
\lambda_{k-1}^{(k-1)}\leftrightarrow \lambda_{k}^{(k)}\label{eq:lambdakk-1}
\end{equation}

Taking the derivative of both sides of \eqref{eq:projection} w.r.t. $\lambda_{n-\!\mu_n\!+1}^{(k-1)}$ and with some operations, we can obtain that
\begin{equation*}
\frac{\mathrm{d}\lambda_{n-\!\mu_n\!+1}^{(k)}}{\mathrm{d}\lambda_{n-\!\mu_n\!+1}^{(k-1)}} =\!
\frac{1+\! \sum\limits_{i=1,z_i\neq 0}^{n-\mu_n}\!\frac{z_i^2}{(\lambda_i^{(k-1)}-\lambda_{n-\mu_n+1}^{(k)})^2}}
{1+\! \sum\limits_{i=1,z_i\neq 0}^{n-\mu_n}\!\frac{(\lambda_{i}^{(k-1)}\!-\!\lambda_{n-\!\mu_n\!+1}^{(k-\!1)})z_i^2}{(\lambda_i^{(k-1)}-\lambda_{n-\mu_n+1}^{(k)})^2}}
\end{equation*}
Since $\lambda_i^{(k-1)}\geq\lambda_{n-\mu_n+1}^{(k-1)}$ for all $i\leq n-\mu_n$, we can obtain that ${\mathrm{d}\lambda_{n-\mu_n+1}^{(k)}}/{\mathrm{d}\lambda_{n-\mu_n+1}^{(k-1)}}> 0$, which implies that $\lambda_{n-\mu_n+1}^{(k)}$ is monotonically strictly increasing w.r.t. $\lambda_{n-\mu_n+1}^{(k-1)}$; therefore,
\begin{equation}
\lambda_{n-\mu_n+1}^{(k-1)}\leftrightarrow \lambda_{n-\mu_n+1}^{(k)} \label{eq:lambda_nn}
\end{equation}

For $n\leq k \leq M$, generally $\mu_n= 1$ and from \eqref{eq:zeta} and \eqref{eq:lambda_nn} we can obtain that
\begin{equation}
\zeta_k \leftrightarrow \lambda_n^{(k)} \leftrightarrow \lambda_n^{(k+1)} \leftrightarrow ... \leftrightarrow \lambda_n^{(M)}  \label{eq:k>n_1}
\end{equation}
 If $\mu_n> 1$, from \eqref{eq:zeta} and \eqref{eq:lambdakk-1_leqn} we can obtain that
\begin{equation}
\zeta_k \leftrightarrow \lambda_{n-\mu_n+1}^{(k)} \leftrightarrow \lambda_{n-\mu_n+2}^{(k+1)} \leftrightarrow ... \leftrightarrow \lambda_n^{(k+\mu_n-1)}  \label{eq:k>n_M}
\end{equation}
Since $\lambda_n^{(k+\mu_n-1)}$ is a simple eigenvalue, considering \eqref{eq:lambda_nn} and \eqref{eq:k>n_M} we can find
\begin{equation}
\zeta_k \leftrightarrow \lambda_n^{(k+\mu_n-1)} \leftrightarrow \lambda_{n}^{(k+\mu_n)}\leftrightarrow ... \leftrightarrow\lambda_n^{(M)} \label{eq:k>nzeta_M}
\end{equation}
From \eqref{eq:k>n_1}-\eqref{eq:k>nzeta_M}, we can obtain that
\begin{equation}
\lambda_n^{(n)}\leftrightarrow \lambda_n^{(M)} \label{eq:lambda_nnnM}
\end{equation}

For $k<n$, the multiplicity of $\lambda_n^{(k-1)}$ w.r.t. $\mathbf{\Psi}_{k-1}$ is $n-k+1$, i.e. $\mu_n=n-k+1$. Then, considering \eqref{eq:zeta}, \eqref{eq:lambdakk-1} and \eqref{eq:lambda_nnnM}, we can obtain that
\begin{equation}
\zeta_k \leftrightarrow \lambda_k^{(k)} \leftrightarrow \lambda_{k+1}^{(k+1)}\leftrightarrow ... \leftrightarrow\lambda_n^{(n)} \leftrightarrow \lambda_n^{(M)} \label{eq:zeta&nn}
\end{equation}

In summary, from \eqref{eq:k>n_1}, \eqref{eq:k>nzeta_M} and \eqref{eq:zeta&nn}, we can conclude that for any $M\geq n$, $\lambda_n^{(M)}$ is monotonically increasing w.r.t. $\zeta_k$ for all $k\leq M$.
\end{IEEEproof}

\bibliographystyle{IEEEtran}
%\bibliography{mybibfile}

\end{document}